\documentstyle[preprint,eqsecnum,aps]{revtex}

\def\go{\mathrel{\raise.3ex\hbox{$>$}\mkern-14mu
             \lower0.6ex\hbox{$\sim$}}}
\def\lo{\mathrel{\raise.3ex\hbox{$<$}\mkern-14mu
             \lower0.6ex\hbox{$\sim$}}}
\def\eps{\varepsilon}
\def\br{{\bf r}}
\def\brp{{\bf r_{\perp}}}

\begin{document}
\draft
\title{
Hydrogen Molecules In a Superstrong Magnetic Field: \\
II. Excitation Levels}

\author{Dong Lai}
\address{Theoretical Astrophysics, 130-33, California Institute of
Technology\\
Pasadena, CA 91125\\
{\rm E-mail: dong@tapir.caltech.edu}}
\author{Edwin E.~Salpeter}
\address{
Center for Radiophysics and Space Research, Cornell University\\
Ithaca, NY 14853\\
\vskip 0.4in}

\date{To be published in {\it Physical Review A}}

\maketitle 
\begin{abstract}

We study the energy levels of H$_2$ molecules in a superstrong 
magnetic field ($B\go 10^{12}$ G), typically found on the surfaces of
neutron stars. The interatomic interaction potentials are calculated
by a Hartree-Fock method with multi-configurations 
assuming electrons are in the ground Landau
state. Both the aligned configurations and arbitrary orientations of
the molecular axis with respect to the magnetic field axis are
considered. Different types of molecular excitations are then studied: 
electronic excitations, aligned (along the magnetic axis) 
vibrational excitations, transverse vibrational excitations 
(a constrained rotation of the molecular axis around the magnetic
field line).
Similar results for the molecular ion H$_2^+$ are also 
obtained and compared with previous variational calculations. 
Both numerical results and analytical fitting formulae are given 
for a wide range of field strengths. 
In contrast to the zero-field case, it is found that the transverse 
vibrational excitation energies can be larger than the aligned 
vibration excitation, and they both can be comparable or larger 
than the electronic excitations. 
For $B\go B_{crit}=4.23\times 10^{13}$ G, the Landau energy of
proton is appreciable and there is some controversy regarding the 
dissociation energy of H$_2$. We show that H$_2$ is bound
even for $B>>B_{crit}$ and that neither proton has a Landau excitation
in the ground molecular state.

\end{abstract}
\bigskip

\widetext

\section{\bf INTRODUCTION}
\label{sec:intro}

Since the pioneering work of Schiff and Snyder\cite{Schiff39}, 
especially during the last 20 years, there has been considerable
interest in the properties of matter in a strong magnetic field. While
the early studies\cite{Elliot60}
were mainly motivated by the fact that high magnetic field
conditions can be mimicked in some semiconductors where a small
effective mass and a large dielectric constant reduce the electric
force relative to the magnetic force, the recent interest in this
problem has been motivated by the huge magnetic field $\sim 10^{12}$ G
already discovered in many neutron stars and the tentative suggestion
for fields as strong as $10^{15}$ G. 
The surface layer of these neutron stars
then consists of highly-magnetized matter. 
Understanding the physical properties
of atoms, molecular chains, and condensed matter in fields of such 
extreme magnitude (see Ref.\cite{Ruderman74} for an early general
review and \cite{Ruder94} for a recent text on atoms in strong
magnetic fields) is important for interpreting the radiation from the
neutron stars that may be observed in the present and future X-ray
satellites (e.g., \cite{Ogelman93}), therefore provides important 
information about the internal structure of neutron stars. 

In superstrong magnetic fields the structure of atoms and condensed
matter is dramatically changed by the fact that the magnetic
force on an electron is stronger than the Coulomb
force it experiences, i.e., the electron cyclotron energy 
(the Landau energy level spacing)
\begin{equation}
\hbar \omega _e = \hbar {e B\over m_ec}
=11.57 B_{12}~{\rm keV},\end{equation}
where $B_{12}$ is the magnetic field strength in units of
$10^{12}~{\rm G}$, is much larger than the typical Coulomb energy. 
In the direction perpendicular to the field, the electrons are
confined to move on cylindrical Landau orbitals around a nucleus.
The orbitals have radii 
\begin{equation}
\rho_m=(2 m+1)^{1/2} {\hat \rho},~~ m=0,1,2,\cdots,
\end{equation}
where ${\hat \rho}$ is the cyclotron radius
\begin{equation}
{\hat \rho}=\biggl({\hbar c\over e B}\biggr)^{1/2}
=a_o\left({B_o\over B}\right)^{1/2}
= 2.57 \times 10^{-10} B_{12} ^{-1/2}~{\rm cm}.
\end{equation}
Here $a_o=\hbar^2/(m_ec^2)$ is the Bohr radius and $B_o$ is the 
atomic unit for the magnetic field strength,
\begin{equation}
B_o={m_e^2e^3 c\over\hbar^3}=2.35 \times 10^9~{\rm G},~~~
b \equiv {B \over B_o}=425\,B_{12}.
\end{equation}
Throughout this paper we consider strong fields in the sense of 
$b>>1$, so that the Coulomb forces act as a perturbation to the 
magnetic forces on the electrons, and 
the electrons are confined to the ground Landau level 
(so called ``adiabatic approximation''\cite{Schiff39}). 
Because of this extreme confinement of electrons in the
transverse direction, the Coulomb force becomes much more effective for 
binding electrons in the parallel direction, therefore giving greatly 
increased binding energy. The atom has a cigar-like structure.
Moreover, it is possible for these elongated
atoms to form molecular chains by covalent bonding along the field 
direction\cite{Ruderman74,Lai92}.

Significant efforts have been devoted to the theoretical
study of atoms in a superstrong magnetic field ($\go 10^{12}$ G)
\cite{Ruder94}. 
The methods that have been employed 
include variational calculations(e.g.,\cite{Rau75}),
Thomas-Fermi-type statistical models\cite{Fushiki92},
density functional theory\cite{Jones85}, and 
self-consistent Hartree-Fock
method\cite{Virtamo76,Neuhauser87,Miller91},
which we consider to be the more theoretically justified and reliable
a method. Accurate calculations of the energy levels of the H atom in
magnetic fields of arbitrary strength have also been
performed\cite{Rosner84}. 
By contrast, there are only limited studies on molecules in
superstrong magnetic field; nearly all of these focus on the molecular
ion H$_2^+$
(\cite{Larsen82,Wunner82,Khersonskii84,LeGuillou84} 
and references therein). As H$_2^+$ is unstable against 
forming H$_2$, understanding the physical properties of H$_2$ 
molecule is of greater practical interest, since H$_2$ is likely to 
exist in the atmosphere of sufficiently cool neutron 
stars\cite{Lai95b,Lai92b}


We have recently calculated the ground state binding energies of
different forms of hydrogen (H, H$^-$, H$_2^+$, 
H$_2$, H$_3$, $\cdots$, H$_{\infty}$)
in a strong magnetic field $B\go 10^{12}$ G ($\!$\cite{Lai92},
hereafter referred as Paper I). In particular, for the first time, 
reliable electronic dissociation energy of H$_2$ molecule in magnetic
field of such magnitude was obtained. In this paper, we extend our
study to consider various excitation levels of the molecule. 

In the zero-field case, to study the molecular spectra, one usually
uses Born-Oppenheimer approximation to separate the motion of the ions 
from that of the electrons. Such a procedure is valid 
if the electronic energy-level spacings are large compared to the
typical energy-level spacings associated with the ion motion. In a
strong magnetic field, however, 
the separation of motion becomes much more 
complicated, even for the hydrogen atom\cite{Lai95a,Avron78,Schmel88}. 
Moreover, as we shall see, in a superstrong magnetic field, the 
energy-level spacings associated with the vibrations of the ions 
can be comparable to or even larger than the spacings of the 
electronic excitations. In this paper, we will use the standard 
Born-Oppenheimer approximation and focus on calculating the
interatomic interaction potential for fixed ion positions (Sec.~III). 
We then obtain the molecular excitation levels based on this 
potential curve (Sec.~IV). 
As in the case of a neutral atom\cite{Lai95a},
it is convenient to define a critical field strength by equating the 
cyclotron energy of the proton $\hbar\omega_p=\hbar(eB/m_pc)$ to the 
typical electronic excitation energy 
($\sim \ln b$ in atomic units), i.e.,
\begin{equation}
b_{crit}\equiv {m_p\over m_e}\ln b_{crit}=1.80\times 10^4;~~~~~
B_{crit}=b_{crit}B_o=4.23\times 10^{13}~{\rm G}.
\end{equation}
We shall give quantitative results for the regime 
$B_o<<B<<B_{crit}$ in Sec.~II-IV, using the standard 
Born-Oppenheimer procedure. Rigorous calculations
for the molecule when $B\go B_{crit}$, 
taking account of the quantum mechanics of the proton motion,
are difficult. Nevertheless, in Sec.~V we shall describe an 
approximate solution to the four-body problem of the H$_2$ molecule 
in the $B>>B_{crit}$ regime, where the effects of
finite proton mass on the electronic states and the energies of the 
molecule are strong, and we give a rigorous lower limit to the 
ground-state dissociation energy.   

Throughout this paper, we shall use nonrelativistic quantum mechanics,
even for extremely strong magnetic field, 
$B\go B_{rel}=(\hbar c/e^2)^2 B_o= 4.414 \times 10^{13}$ G
(note that $B_{rel}$ is close to $B_{crit}$ only by coincidence),
at which the transverse motion of the electron 
becomes relativistic. The nonrelativistic treatment of bound states
is valid for two reasons: 
(i) The energy of a relativistic free electron 
\begin{equation}
E=\left[c^2p_z^2+m_e^2c^4\left(1+2n_LB/B_{rel}\right)\right]^{1/2},
\end{equation}
where $p_z$ is the linear momentum along the field axis, $n_L$ 
is the quantum number for the Landau excitations, reduces to 
$E\simeq m_ec^2+p_z^2/(2m_e)$ as along as 
the electron remains in the ground Landau level and 
nonrelativistic in the $z$-direction;
(ii) The shape of the Landau wavefunction in the relativistic theory 
is the same as in the nonrelativistic theory
(as we see from the fact that $\hat\rho$ 
is independent of mass). Therefore, 
as long as $E_B/(m_e c^2)<<1$, where
$E_B$ is the binding energy of the bound state, the relativistic
effect remains a small correction\cite{Angelie78}.

The paper is organized as follows. In Sec.~II we consider some general
features and approximate scaling relations for various excitation
levels. Sec.~III contains a detailed description of our method for
calculating the interatomic interaction potential. The numerical
results and 
fitting formulae for the molecular excitation levels are presented 
in Sec.~IV. We study the electronic structure of the molecule 
in the $B>>B_{crit}$ regime and consider the effects of finite 
proton mass on the energies in Sec.~V. Our general conclusion 
is presented in Sec.~VI. 
Appendix A summarizes some useful mathematical relations for the 
Coulomb integrals of Landau functions, and in Appendix B we discuss
a refined method for calculating the electronic energy
of H$_2^+$ for general orientation of the molecular axis. 

\section{\bf QUALITATIVE DISCUSSION AND APPROXIMATE SCALING RELATIONS 
FOR EXCITATION ENERGIES}
\label{sec:app}

In a superstrong magnetic field satisfying $b>>1$,
the spectra of a single H atom can be specified by two quantum numbers
$(m,\nu)$, where $m$ measures the mean transverse distance (Eq.~[1.2])
of the electron to the proton, while $\nu$ is the number of nodes
of electron's $z$-wavefunction (along the field direction). 
The wavefunction of the $(m,\nu)$ state in cylindrical
coordinates $(\rho, \phi, z)$ is given by
\begin{equation}
\Phi_{m\nu}=W_m(\brp)f_{m\nu}(z),\end{equation}
where $W_m$ is the ground state Landau wavefunction
\begin{equation}
W_m(\brp)\equiv W_m(\rho,\phi)={1 \over \hat\rho\sqrt {2 \pi m!}} 
\biggl({\rho \over \hat\rho\sqrt 2}
\biggr )^m e^{-\rho ^2 /4\hat\rho^2} e^{-i m \phi}.\end{equation}
The states with $\nu \neq 0$ resemble a zero-field hydrogen atom
with small binding energy $|E_\nu|\simeq 1/(2\nu^2)$\cite{Haines69}
and we shall mostly focus on the tightly-bound states with $\nu=0$.
For the ground state $(0,0)$, the sizes $L_\perp$ and $L_z$ of the
atomic wavefunction perpendicular and parallel to the field and the 
binding energy $|E_a|$ (in atomic units) are given by
\begin{equation}
L_\perp\sim \hat\rho={1\over b^{1/2}},~~~~
L_z\sim {1\over l},~~~~
|E_a|\simeq 0.16\,l^2;~~~~~~~l\equiv \ln b.
\end{equation}
For the tightly-bound excited states $(m,0)$ we have 
similar relations but with
$\hat\rho$ replaced by $[(2m+1)/b]^{1/2}$ and $l$ replaced by 
$l_m\equiv \ln [b/(2m+1)]$, so that 
\begin{equation}
E_a(m)\simeq -0.16\,l_m^2.
\end{equation}
Recall that in atomic units, mass is in units of the electron mass
$m_e$, energy is expressed in units of
$e^2/a_0 = 2~{\rm Ry}$, length is in units of Bohr radius $a_0$,
and the units for magnetic field is $B_o$ (Eq.~[1.4]).
The numerical factor $0.16$ in Eqs.~(2.3)-(2.4) is an approximate
value for $B_{12}\go 1$. For convenience, accurate numerical 
results for $E_a(m)$ are listed in Table I.
\footnote{A more accurate fitting formula for the ground state
binding energy of H atom is 
$|E_a|=0.16\,A\,l^2$, with
$$A=\cases{1+1.36\times 10^{-2}\,[\ln (1000/b)]^{2.5},
	& if $b<10^3$;\cr
    1+1.07\times 10^{-2}\,[\ln (b/1000)]^{1.6}, & if $b\ge 10^3$.
\cr}$$
} 

In a superstrong magnetic field, the mechanism of forming 
molecules is quite different from the zero-field case
(Paper I,\cite{Ruderman74}). The spins of the electrons of the 
atoms in a strong magnetic field are all aligned anti-parallel to 
the magnetic field, and therefore two atoms in their 
ground states do not easily bind together
according to the exclusion principle. Thus two H atoms,
both in the $m=0$ ground state, do not form tightly bound molecule. 
Instead, one H atom has to be excited to the $m=1$ state. The 
two H atoms, one in the ground state ($m=0$), another 
in the $m=1$ state then form the ground state of H$_2$ molecule by 
covalent bonding. Since the ``activation energy'' for exciting 
an electron in the H atom from Landau orbital $m$ to $(m+1)$
is small (see Eq.~[2.4]), the resulting molecule is stable.
The interatomic separation $Z_o$ and the dissociation energy $D$ of the
H$_2$ molecule scale approximately as 
\begin{equation}
Z_o=\xi L_z\sim {\xi\over l},~~~~~
D\sim {l\over Z_o}\sim {l^2\over\xi},
\end{equation}
where the dimensionless factor $\xi$ decreases very slowly with
increasing $B$ (e.g., $\xi\simeq 2.0$ for $B_{12}=0.1$ and $\xi\simeq
0.75$ for $B_{12}=100$; see Table I of Paper I and our Eq.~[5.2]). 

Another mechanism of forming H$_2$ molecule in a superstrong magnetic 
field is to let both electrons occupy the same $m=0$ Landau state,
while one of them occupies the 
$\nu=0$ orbital and another $\nu=1$ orbital.
This costs no ``activation energy''. However, the resulting molecule 
tends to have small dissociation energy, of order a Rydberg.
We shall refer to this electronic state of the molecule as the 
{\it weakly-bound state}, to the states formed by two electrons
in the $\nu=0$ orbitals as the {\it tightly-bound states}.
As we will see below, as long as $l>>1$, the weakly-bound state 
only constitutes an excited energy level of the molecule.
\footnote{In several recent papers\cite{Korolev94} on 
the molecular binding in strong magnetic field, Korolev and Liberman 
failed to identify the tightly-bound states. Also, their
variational calculation of the weakly-bound state significantly 
underestimates the binding energy because it neglects the 
overlapping of the electron wavefunctions. As a result, their
claim that hydrogenlike gas in strong magnetic field can form 
Bose-Einstein condensate is incorrect 
(see also\cite{Lai95c,Ortiz95}).}

We now consider various molecular excitations and derive approximate
scaling relations for the excitation energies.

\subsection{\bf Electronic Excitations}

The electronic excitations of H$_2^+$ are similar to those of
the H atom, namely the electron can occupy different $m$ Landau orbitals. 
Thus $m=0$ is the ground state, $m=1,2,\cdots$ are the excited states
(although they are not necessarily bound relative to the free atom 
in the ground state). 

There are two types of electronic excitations in H$_2$. 
(i) The electrons can occupy different orbitals 
other than the ground state $(m_1,m_2)=(0,1)$, 
giving rise to the tightly-bound ($\nu=0$) electronic excitations. 
For example, the first excited level is $(0,2)$, the
second excited level is $(0,3)$, etc.
The number of single $m$-excitation states $(m_1,m_2)=(0,m_2)$ which 
are bound relative to two isolated H atoms in the ground state
is expected to increase as the magnetic field increases.
Double $m$-excitations are also possible,
but as we shall see, they are bound only when the magnetic 
field strength is much higher than $10^{13}$ G.
The energy spacing between the two adjacent 
electronic states $(0,m)$ and $(0,m+1)$ is 
\begin{equation}
\Delta E_m \sim l\,\ln\biggl({2m+3\over 2m+1}\biggr).
\end{equation} 
Thus as $m$ increases, the energy spacing decreases. 
(ii) The molecule is formed by two electrons
in the $(m,\nu)=(0,0)$ and $(0,1)$ orbitals. The dissociation energy of 
this weakly-bound state is of order a Rydberg, and does not depend
sensitively on the magnetic field strength. Note that for relatively 
small magnetic field ($B_{12}\lo 0.2$), the weakly-bound state
actually has lower energy than the tightly-bound states (see Sec.~IV.A),
i.e., $b\go 10^2$ is required for the
``strong field'' regime to apply fully. 

\subsection{\bf Aligned Vibrations}

In the Born-Oppenheimer approximation, the motion of the two protons
is governed by the interatomic potential $U(Z,R_{\perp})$, i.e., 
the electronic energy when the relative positions of
the protons are kept at $Z$ along the field direction and $R_{\perp}$
perpendicular to it. We first consider the aligned vibrational 
excitations for oscillations of $Z$ about the equilibrium separation
$Z_o$. For this purpose we need to estimate the excess potential 
$\delta U(\delta Z)\equiv U(Z_o+\delta Z,0)-U(Z_o,0)$.

Since $Z_o$ is the equilibrium position, the sum of the 
first order terms in $\delta Z$,
coming from proton-proton, electron-electron,
proton-electron Coulomb energies and quantum mechanical electron
kinetic energy, must cancel. Thus we have 
$\delta U\propto (\delta Z)^2$ for small $\delta Z$. 
Consider various contributions to the energy of the molecule: 
The proton-proton interaction is $1/Z$ (in atomic units) without a 
logarithmic factor; but the dominant contribution 
is the proton-electron Coulomb energy
$\sim l/Z$, where the logarithmic factor $l>>1$ comes from the Coulomb 
integral over the ``cigar-shaped'' electron distribution. Both $l$ and
$Z^{-1}$ change as $Z_o\rightarrow Z_o+\delta Z$, but the largest
change comes from 
the quadratic term $\delta(Z^{-1})\sim (\delta Z)^2/Z_o^3$. Thus the 
excess potential is of order 
\begin{equation}
\delta U(\delta Z)\sim l\,{(\delta Z)^2\over Z_o^3}
\sim \left(\xi^{-3}l^4\right)\,(\delta Z)^2.
\end{equation}

In atomic (electron) units the reduced mass of the proton-pair in
H$_2$ is $\mu={m_p/(2m_e)}$,
where $m_p$ and $m_e$ are proton and electron mass (for HD the factor
$1/2$ is replaced by $2/3$). For small-amplitude oscillations in the
potential of equation (2.7), we obtain a harmonic oscillation spectrum 
with excitation energy quanta $\hbar \omega_{\parallel}$ given by 
\begin{equation}
\hbar \omega_{\parallel} \sim
\xi^{-3/2}\,l^2\,\mu^{-1/2},\end{equation}
for a molecule in the ground electronic state.
The scaling with $B$ of $\hbar \omega_{\parallel}$ is thus almost 
the same as the dissociation energy $D$ in Eq.~(2.5). The number 
of aligned vibrational levels is $n_{\parallel max}\sim
D/\hbar\omega_\parallel\sim (\xi\mu)^{1/2}$, where $\xi$ decreases
even more slowly with increasing field strength than $l^{-1}$ does. 

\subsection{\bf Transverse Vibrations}

The strong magnetic field breaks the rotational symmetry
for the molecular axis and, instead of rotations of the field-free case
we have oscillations in the two-dimensional plane of 
the ${\bf R_\perp}$ vector\footnote{
Strictly speaking, the transverse vibration and the aligned 
vibration are coupled, and they are governed by the two dimensional
potential $U(Z,R_{\perp})$. Since the transverse vibrational
excitation energy is larger than the aligned vibrational excitation, 
the timescale for the protons to adjust their
$Z-$positions is much longer than the timescale for oscillations with 
$R_{\perp}\neq 0$ and we can consider transverse vibrations with 
fixed values of $Z$. However, since $\delta Z<<Z$, an 
approximate separation is possible with $Z$ replaced by $Z_o$
for the transverse vibrations.}.
The degeneracy in the azimuthal angle $\phi$ is still retained. 
To study the transverse vibration spectrum, we 
need to estimate the order of magnitude of the excess potential 
$\delta U(R_\perp)\equiv U(Z_o,R_\perp)-U(Z_o,0)$. 

As mentioned before, the factor $l$ in the expression $l/Z_o$ for the
dissociation energy $D$ (Eq.~[2.5]) 
comes from a Coulomb integral over the electron charge
distribution. This integral is of the form $\ln (L_z/\hat\rho)$, where 
$\hat\rho=b^{-1/2}$ is the typical size of the electron wavefunction
perpendicular to the field for $R_\perp=0$. 
When the protons are displaced by $R_\perp$ from the 
electron distribution axis, the Coulomb integral can be approximately 
obtained by replacing $\hat\rho$ with
$(\hat\rho^2+R_\perp^2)^{1/2}$. Our order of magnitude expression for 
$\delta U$ is then 
\begin{equation}
\delta U(R_{\perp})\sim {1\over
2Z_o}\ln\left(1+\hat\rho^{-2}R_\perp^2\right)
\sim \xi^{-1}l\,\ln\left(1+b\,R_\perp^2\right).
\end{equation}
Equation (2.9) holds for any $R_\perp<<Z_o\sim \xi\,l^{-1}$, but it
can be approximated by a quadratic expression for the small-amplitude
case of $R_\perp\lo \hat\rho=b^{-1/2}<<Z_o$. In this approximation
we have $\delta U\sim \xi^{-1}l\,b\,R_\perp^2$. The energy quanta for
the small-amplitude transverse vibration is then
\begin{equation}
\hbar \omega_{\perp 0}\sim \left(\xi^{-1}\,l\,b\right)^{1/2}\mu^{-1/2},
\end{equation}
where the subscript $0$ indicates that we are at the moment neglecting 
the magnetic forces on the protons which, in the absence of Coulomb
forces, lead to the cyclotron motions of the protons. 
Note that $\hbar\omega_{\perp 0}$ in Eq.~(2.10) increases as $b^{1/2}$ 
with increasing field strength, faster than the logarithmic behavior of 
$\hbar\omega_\parallel$ and $D$, but slower than
the linear behavior of the cyclotron energy. 
For sufficiently large $b>>1$ we have 
$\hbar\omega_{\perp 0}>>\hbar\omega_\parallel$. 
However, the quadratic harmonic oscillator approximation 
is valid only for $R_\perp^2$
up to $\sim \hat\rho^2=b^{-1}$, i.e., for $\delta U$ only up to 
$\delta U_{ho}\sim \xi^{-1}l$, which is less than the maximum 
possible potential $\Delta U_{max}\sim D\sim \xi^{-1}l^2$. 
The number of harmonic oscillation levels in the quadratic regime is
then 
\begin{equation}
n_{\perp ho}\sim {\delta U_{ho}\over\hbar\omega_{\perp 0}}
\sim \xi^{-1/2}\left({\mu\,l\over b}\right)^{1/2}
\sim \left({b_{crit}\over b}\right)^{1/2}.
\end{equation}
The degeneracy of the $n_{\perp}$-th harmonic oscillation level is 
$n_{\perp}$. For $n_{\perp ho}>>1$, the statistical weight 
of all harmonic oscillation levels is of order $(n_{\perp ho})^2$. 
If we neglect the difference between $\xi$ and unity
(and between $\mu$ and $m_p/m_e$), we see that 
$n_{\perp ho}$ would be less than unity when $B\go B_{crit}$, where
$B_{crit}$ is defined in Eq.~(1.5). 

We now consider large amplitude transverse oscillations assuming that
the magnetic force on the proton can be neglected. 
For a transverse oscillation wavefunction where the maximum value
$R_{max}$ of $R_\perp$ (the outer classical turning point) 
satisfies $\hat\rho\lo R_{max}\lo Z_o$, we must use the logarithmic
form of Eq.~(2.9) for the potential $\delta U(R_\perp)$. The 
energy level-spacing decreases with increasing $R_{max}$. 
We can calculate the number of nodes $n_\perp(R_{max})$ of the 
wavefunction as a function of $R_{max}$ from a WKB integral of
the wave number $k(R_\perp)$ over $dR_\perp$. Since we only need an 
order of magnitude estimate, we replace the integral by 
$k(R_{max})R_{max}$, where $k(R_\perp)\sim [\mu\delta U(R_\perp)]^{1/2}$. 
Using Eq.~(2.9) this gives
\begin{equation}
n_\perp(R_{max})\sim \left[\mu\,\xi^{-1}l\,\ln(1+bR_{max}^2)\right]^{1/2}
R_{max}.
\end{equation}
The maximum number of nodes $n_{\perp max}$ can be obtained by
substituting $Z_o\sim \xi/l$ for $R_{max}$.  Neglecting $\ln l$ 
compared with $l$ itself, we have
$n_{\perp max}\sim (\xi\mu)^{1/2}$, independent of field strength and
the same order of magnitude as $n_{\parallel max}$. 

Because of the azimuthal symmetry in the two-dimensional 
${\bf R_\perp}$-plane,
the total statistical weight of the transverse excitation levels is 
$\sim n_{\perp max}^2\sim \xi\mu$. If $b>>b_{crit}$, 
$n_{\perp ho}$ in Eq.~(2.11) would be much less than unity
and the zero-point energy $\eps_{\perp zp}$, i.e., 
the spacing between the lowest levels, is not given by Eq.~(2.10). 
Formally, one could use Eq.~(2.9) and estimate the zero-point 
vibration amplitude as the value of $R_{max}$ for which Eq.~(2.12)
gives $n_\perp=1$. This would give a zero-point energy which 
is less than $D$, but this expression is incorrect, since 
the neglect of the magnetic forces on the protons 
is unjustified when $B>>B_{crit}$. The cyclotron energy of the proton
is $\hbar\omega_p=\hbar eB/(m_pc)=(m_e/m_p)\,b$ (a.u.).
The ratio $\omega_p/\omega_{\perp 0}$ is of order
$(b\,m_e/l\,m_p)^{1/2}=(b/b_{crit})^{1/2}$ (omitting the factor $\xi$).
When $\hbar\omega_p$ is much larger than $\hbar\omega_{\perp 0}$, 
the magnetic forces on the protons are important. We will return
to this subtle issue in Sec.~V. 


\section{\bf METHODS FOR CALCULATING THE INTERATOMIC POTENTIAL}

In the Born-Oppenheimer approximation, the interatomic potential
$U(Z,R_{\perp})$ is given by the total 
electronic energy $E(Z,R_{\perp})$ of the system when the relative
positions of the protons are $Z$ along the field direction and
$R_{\perp}$ perpendicular to it.
Once $E(Z,R_{\perp})$ is obtained, the electronic equilibrium state
can also be determined by locating the minimum of the $E(Z,0)$ curve. 

\subsection{\bf The Aligned Case: $R_{\perp}=0$}

Our method for calculating $E(Z,0)$ is the same as in Paper I. It
can also be used to obtain the energy curves for the excited
electronic states. Here we summarize and extend our method to take
account of ``configuration interaction'' in H$_2$ more accurately. 

\subsubsection{H$_2^+$ Molecular Ion}

For H$_2^+$, the Hamiltonian for the electron is
\begin{equation}
H_o=H_B-{\hbar^2 \over 2 m_e}{\partial^2\over\partial z^2}
-{e^2 \over r_{A}} -{e^2 \over r_{B}},\end{equation}
where $r_A$ and $r_B$ are the distances between the electron and the
two fixed protons, located at $z=\pm Z/2$ along $z$-axis.
In Eq.~(3.1), $H_B$ is the magnetic part of the Hamiltonian
\begin{equation}
H_B={1\over 2m_e}\biggl({\bf p_{\perp}}
+{e\over c}{\bf A}\biggr)^2+{e\over m_ec}{\bf B}\cdot {\bf S},
\end{equation}
where ${\bf A}={\bf B}\times\br/2$ and ${\bf S}$ is the electron spin
operator. Note that for electrons in the ground Landau level, we have
\begin{equation}
H_B [W_m(\brp)\chi(\downarrow)]=0,\end{equation}
where $\chi(\downarrow)$ is the electron spinor with the spin aligned 
in the $-z$-direction (anti-parallel to the field). Thus we can set
$H_B=0$. With the electron wavefunction given by $\Phi _{m0}({\bf r})=
W_m(\brp) f_{m0}(z)$, we average over the transverse direction 
and obtain a one-dimensional Schr\"odinger equation
\begin{equation}
-{ \hbar^2 \over 2 m_e {\hat \rho}^2}{d^2\over
dz^2}f_{m0}-{e^2\over\hat\rho}
\tilde V_m(z)f_{m0}=\eps_{m0} f_{m0}.
\end{equation}
Here the averaged potential is given by
\begin{equation}
\tilde V_m(z)=\int \! d^2 \brp |W_m (\brp)|^2
\biggl({1 \over r_A}+{1 \over r_B}\biggr)
=V_m\biggl(z-{Z\over 2}\biggr)+V_m\biggl(z+{Z\over 2}\biggr),
\end{equation}
where 
\begin{equation}
V_m(z)\equiv\int \! d^2 \brp |W_m (\brp)|^2 
{1 \over r},\end{equation}
which can be evaluated numerically (Paper I). In Eqs.~(3.4)-(3.6) and 
hereforth we employ $\hat\rho$ as the length unit in all wavefunctions
and average potentials (except otherwise noted). We solve the 
eigenvalue $\eps_{m0}$ by integrating Eq.~(3.4) numerically from
$z=+\infty$ to $z=0$ subject to appropriate boundary conditions (Paper
I). The total electronic energy is then given by
\begin{equation}
E(Z,0)=\eps_{m0}+{e^2\over Z}.\end{equation}
Clearly, $m=0$ is the ground state, while $m=1,2,\cdots$ are the
excited electronic states. 

We also note that the excited state of H$_2^+$ 
in which the electron occupies the $\nu>0$ orbital is not bound
relative to the free atom in the ground state.

\subsubsection{H$_2$ Molecule: Tightly-Bound States
$(m,\nu)=(m_1,0),(m_2,0)$}

For H$_2$, we use the Hartree-Fock (HF) method to take account of the
interaction between the electrons. The Hamiltonian of the system is
\begin{equation}
H=H_o(1)+H_o(2)+{e^2 \over r_{12}} +{e^2 \over Z},\end{equation}
where $H_o$ is given by Eq.~(3.1) and $r_{12}\equiv |{\bf r_1}
-{\bf r_2}|$. For the $(m_1,m_2)$ electronic
state ($m_1\neq m_2$), the two basis wavefunctions (orbitals) for the
electrons are 
\begin{eqnarray}
\Phi_{m_10}(\br) &=& W_{m_1}(\brp)f_{m_10}(z),\\
\Phi_{m_20}(\br) &=& W_{m_2}(\brp)f_{m_20}(z).
\end{eqnarray}
The two-electron wavefunction is then given by
\begin{equation}
\Psi (\br_1,\br_2)={1 \over \sqrt 2}
[\Phi_{m_10}(\br_1)\Phi_{m_20}(\br_2)
-\Phi_{m_10}(\br_2)\Phi_{m_20}(\br_1)].\end{equation}
After averaging over the transverse direction, the standard HF
equations reduce to a set of one-dimensional equations for $f_{m_10}$
and $f_{m_20}$:
\begin{equation}
\biggl[-{\hbar^2\over 2m_e\hat\rho^2}{d^2\over dz^2}-{e^2\over\hat\rho}
\tilde V_m(z)+{e^2\over\hat\rho}K_m(z)-\eps_m\biggr]f_{m0}(z)=
{e^2\over\hat\rho}J_m(z),~~~m=m_1,~m_2,
\end{equation}
where $\tilde V_m$ is given by Eq.~(3.5); the direct and 
exchange potentials $K$ and $J$ are given by
\begin{eqnarray}
K_{m_1}(z) &=& \int\!dz'f_{m_20}(z')^2 D_{m_1m_2}(z-z'), \\
J_{m_1}(z) &=& f_{m_20}(z)\int\!dz'
f_{m_10}(z')f_{m_20}(z')E_{m_1m_2}(z-z'),
\end{eqnarray}
and similarly for $K_{m_2}$ and $J_{m_2}$. In Eqs.~(3.13)-(3.14),
$D_{m_1m_2}$ and $E_{m_1m_2}$ are the direct and exchange 
interaction kernels defined by
\begin{eqnarray}
D_{m_1m_2}(z_1-z_2) &=& \int\! d^2\br_{1\perp}d^2\br_{2 \perp}
|W_{m_1}(\br_{1\perp})|^2|W_{m_2}(\br_{2\perp})|^2{1 \over r_{12}},\\
E_{m_1m_2}(z_1-z_2) &=& \int\! d^2\br_{1\perp}d^2\br_{2\perp}
W_{m_1}(\br_{1\perp})W_{m_2}(\br_{2\perp})
W_{m_1}^*(\br_{2\perp})W_{m_2}^*(\br_{1\perp}){1\over r_{12}}.
\end{eqnarray}
The functions $D_{m_1m_2}(z)$ and $E_{m_1m_2}(z)$ are
related to the Coulomb interaction potential $V_m$ (Eq.~[3.6]) by
\begin{eqnarray}
D_{m_1m_2}(z) &=&\sum_{s=0}^{m_1+m_2}d_s(m_1,m_2){1\over\sqrt{2}}
V_s\biggl({z\over\sqrt{2}}\biggr),\\
E_{m_1m_2}(z) &=&\sum_{s=0}^{m_1+m_2}e_s(m_1,m_2){1\over\sqrt{2}}
V_s\biggl({z\over\sqrt{2}}\biggr),
\end{eqnarray}
where the coefficients $d_s$ and $e_s$ are given in Paper I.
We solve Eq.~(3.12) numerically using a shooting algorithm
(for detail, see Paper I). Once the wavefunction $f_{m0}(z)$ and
the eigenvalues $\eps_{m0}$ are obtained, the
total electronic energy of the system is calculated via
\begin{eqnarray}
E &=& \langle\Psi |H|\Psi\rangle \nonumber\\
&=&{e^2\over Z}+\eps_{m_10}+\eps_{m_20}-
{e^2\over\hat\rho}\int\!\! dz_1dz_2f_{m_10}(z_1)^2f_{m_20}(z_2)^2
D_{m_1m_2}(z_1-z_2)\nonumber\\
&+& {e^2\over\hat\rho}
\int\!\! dz_1dz_2f_{m_10}(z_1)f_{m_20}(z_2)f_{m_10}(z_2)f_{m_20}(z_1)
E_{m_1m_2}(z_1-z_2),
\end{eqnarray}
where the $4$th term on the right hand side represents the electron 
direct interaction $(-E^{dir})$, and the $5$th term the exchange 
interaction $(-E^{exch})$. 

The Hartree-Fock method discussed above can be used to obtain
accurately the electronic energy near the equilibrium separation $Z_o$.
However, as noted in Paper I, as $Z$ increases, the resulting $E(Z,0)$
becomes less reliable. Moreover, as $Z\rightarrow\infty$, $E(Z,0)$
does {\it not} approach the sum of the energies of two isolated atoms,
one in the $m_1$th state, another in the $m_2$th state. The reason is
that as $Z$ increases, a second configuration of electron orbitals 
becomes more and more degenerate with the first 
configuration in Eq.~(3.11), and there must be 
mixing of these two different configurations.
This ``configuration interaction'' also occurs in the zero-field
H$_2$ molecule\cite{Slater63}. Here
the electron configuration that mixes with $\Psi_1\equiv \Psi$
(Eq.~[3.11]) is 
\begin{equation}
\Psi_2 (\br_1,\br_2)={1 \over \sqrt 2}
[\Phi_{m_11}(\br_1)\Phi_{m_21}(\br_2)
-\Phi_{m_11}(\br_2)\Phi_{m_21}(\br_1)],\end{equation}
which is the same as $\Psi_1$ except $\nu=1$ in the electron orbitals. 
Both $\Psi_1$ and $\Psi_2$ have the 
same symmetry with respect to the Hamiltonian in Eq.~[3.8]:
the total angular momentum along the $z$-axis is $M_{Lz}=1$, the 
total electron spin is $M_{Sz}=-1$, and both $\Psi_1$ and 
$\Psi_2$ are even with respect to 
the operation $\br_i\rightarrow -\br_i$. 
As a result, the matrix element $\langle \Psi_1 |H |\Psi_2 \rangle$ 
is nonzero.

To take account of the mixing of these two 
configurations, we need to extend the standard 
HF method involving one configuration to 
{\it HF with multi-configurations} (HFMC).
This is done as follows. We calculate the energies and wavefunctions
for both $\Psi_1$ and $\Psi_2$ using the HF equations 
(Eq.~[3.12] with $\nu=0$ and $\nu=1$). 
The matrix elements $H_{ij}$ ($i,j=1,2$) are then calculated:
$H_{11}=\langle \Psi_1 |H |\Psi_1 \rangle$ is given by Eq.~(3.19) and 
the expression for $H_{22}=\langle \Psi_2 |H |\Psi_2 \rangle$ 
is similar. The mixing matrix element is given by
\begin{eqnarray}
H_{12} &=&\langle \Psi_1 |H |\Psi_2 \rangle \nonumber\\
&=& {e^2\over\hat\rho}\int\!\! dz_1dz_2 f_{m_10}(z_1)f_{m_11}(z_1)
f_{m_20}(z_2)f_{m_21}(z_2)D_{m_1m_2}(z_1-z_2)\nonumber\\
&-&{e^2\over\hat\rho}
\int\!\! dz_1dz_2 f_{m_10}(z_1)f_{m_21}(z_1)f_{m_20}(z_2)f_{m_11}(z_2)
E_{m_1m_2}(z_1-z_2).
\end{eqnarray}
The total electronic energy in this HFMC scheme is obtained by 
solving the secular equation
$\det |H_{ij}-E \delta_{ij}| =0$, which yields, for the lowest energy
state
\begin{equation}
E={1 \over 2} (H_{11}+H_{22})-{1 \over 2} [(H_{11}-H_{22})^2
+4 H_{12}^2]^{1/2}.
\end{equation}

In Fig.~1, we show the electronic energy curves of H$_2$
at $B_{12}=1$, obtained using our HFMC method. The 
tightly-bound electronic states
are $(m_1,m_2)=(0,1),~(0,2)$ and $(0,3)$. These are the only states
for which the minimum in the energy curves are less than the energy 
$2E_a=-323$ eV of two isolated atoms in the ground state. 
Notice that as $Z$ increases,
the molecular electronic energy becomes larger than $2E_a$, 
reflecting the fact that, in a superstrong magnetic field, 
forming such a tightly-bound molecule requires first 
activating one of the atoms to an excited 
state. However, as $Z$ increases, the energy of the
$(m_1,m_2)$ state does approach $E_a(m_1)+E_a(m_2)$.  
Near the equilibrium separation, the energy obtained using HFMC agrees
well with that of standard HF (the difference is 
less than $1\%$). Thus the standard HF is adequate for determining
the equilibrium electronic energy of the molecule. However,
the HFMC method is crucial to obtaining the correct
large $Z$ behavior of $E(Z,0)$, therefore the aligned vibrational
energy levels of the molecule (Sec.~IV.B).

\subsubsection{H$_2$ Molecule: Weakly-Bound State $(m,\nu)=(0,0),(0,1)$}

These states can be similarly calculated using the HF method. Instead of
Eqs.~(3.9)-(3.10), the electron orbitals are $\Phi_{00}$ and $\Phi_{01}$. 
Fig.~2 shows an example of the energy curve at $B_{12}=1$. 
Clearly, the $E(Z,0)$ curve of such state is much shallower than those
of the tightly-bound states discussed in Sec.~III.A.2. 
In the limit of $Z\rightarrow\infty$, the energy curve approachs
$2E_a$, i.e., no ``activation energy'' is needed to form a molecule
in the weakly-bound state.

\subsection{\bf General Molecular Axis Orientation: $R_{\perp}\neq 0$}

Unlike the case of Sec.~III.A when the molecular axis coincides 
with the magnetic field direction, where we can obtain the 
interatomic potential $E(Z,0)$ with great accuracy,
in the case when the molecular axis deviates from the magnetic field 
direction, the electronic energy $E(Z,R_{\perp})$ is much harder to
calculate. This is because the azimuthal symmetry of the transverse
wavefunction of an electron is broken. Although the electrons still
stay in the ground Landau level, $m$ in the Landau wavefunction 
$W_m(\brp)$ (Eq.~[2.2]) is no longer a 
good quantum number, and the transverse wavefunction
of an electron must involve mixing of different $m$-states. 
Nevertheless, we can still obtain a reasonable 
upper limit for the interatomic potential curve $E(Z,R_{\perp})$,
and hence an upper limit for the transverse vibrational excitation 
energy quanta $\hbar\omega_{\perp 0}$. 
We consider two ansatzs, appropriate for small $R_{\perp}$ and large
$R_{\perp}$ respectively. 

\subsubsection{Ansatz A}

Suppose the two protons are located at $(\pm R_{\perp}/2,0,\pm Z/2)$
in a rectangular coordinate system. 
For sufficiently small $R_{\perp}$, 
the transverse wavefunction is expected 
to be close to $W_m(\brp)$. Thus we assume
the electron wavefunction in H$_2^+$ is given by $\Phi_{m0}({\bf r})= 
W_m(\brp)f_{m0}(z)$. The equation for $f_{m0}$ 
is the same as Eq.~(3.4), except that the potential 
$\tilde V_m(z)$ is replaced by 
\begin{equation}
\tilde V_{mm}(z,R_{\perp}/2)=
V_{mm}\biggl(|z-{Z\over 2}|,{R_{\perp}\over 2}\biggr)
+V_{mm}\biggl(|z+{Z\over 2}|,{R_{\perp}\over 2}\biggr),
\end{equation}
where 
\begin{eqnarray}
V_{mm}(z,R_{\perp}/2) &\equiv & 
\int \! d^2 \brp |W_m (\brp)|^2
{1 \over |\br-{\bf R_{\perp}}/2|}\nonumber\\
&=& \int_0^{\infty}\!\!dq\exp\biggl(-{1\over 2}q^2-q|z|\biggr)
J_0\biggl({qR_{\perp}\over 2}\biggr)L_m\biggl({1\over 2}q^2\biggr),
\end{eqnarray}
(see Appendix A). Here $J_0$ is the Bassel function of zeroth order and 
$L_m$ is the Laguerre polynormial of order $m$ \cite{Abramowitz72}.
We use a standard quadrature algorithm (e.g., \cite{Press87})
to evaluate Eq.~(3.24). 
The Schr\"odinger equation similar to Eq.~(3.4) can be solved to
determine the eigenvalue $\eps_{m0}(Z,R_{\perp})$, and the total
electronic energy is then given by
\begin{equation}
E_{m0}(Z,R_{\perp})=\eps_{m0}(Z,R_{\perp})
+{e^2\over (Z^2+R_{\perp}^2)^{1/2}}.\end{equation}
As noted before, in this general situation, $m$ is not a good
quantum number, but we nevertheless use it to distinguish different
electronic state.

In this ansatz, the equations for H$_2$ are also similar to those
in Sec.~III.A. We still assume the electron orbitals to be
given by Eqs.~(3.9)-(3.10). The HF equations (3.12)-(3.14) remain
valid except the ion-electron interaction potential $\tilde V_m(z)$ is
replaced by $\tilde V_{mm}(z,R_{\perp}/2)$. The electron-electron
interaction kernels are unchanged. The total electronic energy is
still given by Eq.~(3.19) with $e^2/Z$ replaced by
$e^2/(Z^2+R_{\perp}^2)^{1/2}$.

We now estimate the regime of validity of this ansatz. 
As an example, let us consider the ground electronic state of H$_2^+$.
In general, the transverse wavefunction of the electron is
a superposition of different Landau ground state wavefunctions, i.e.,
\begin{equation}
\Phi_{\perp}(\brp)=\sum_m A_mW_m(\brp),
\end{equation}
and $\Phi(\br)=\Phi_{\perp}(\brp)f(z)$ is the total wavefunction
(see also Appendix B).
For simplicity, just consider the first two terms in the expansion
(3.26), i.e., $\Phi_{\perp}(\brp)=A_oW_o(\brp)
+A_1W_1(\brp)$, with $|A_1|\ll |A_o|$ for the ground state. 
Substitute $\Phi(\br)$ into the Schr\"odinger equation and average
over $\brp$, we obtain (in atomic units)
\begin{eqnarray}
-{1\over 2}{d^2\over dz^2}f-\tilde V_{00}(z)f+{A_1\over A_o}\tilde
V_{01}(z)f 
&=&\eps f,\\
-{1\over 2}{d^2\over dz^2}f-\tilde V_{11}(z)f+{A_o\over A_1}\tilde
V_{10}(z)f 
&=&\eps f,
\end{eqnarray}
where $\tilde V_{mm'}$ is defined similar to Eq.~(3.24).
Since $|\tilde V_{01}|\ll |\tilde V_{00}|$ and
$|\tilde V_{01}|\ll |\tilde V_{11}|$, from Eqs.~(3.27)-(3.28) we have
$A_1/A_o\simeq\tilde V_{10}/(\tilde V_{11}-\tilde V_{00})$.
Substitute this into the Eq.~(3.27), we have
\begin{equation}
-{1\over 2}{d^2\over dz^2}f-\tilde V_{00} f+
{\tilde V_{01}\tilde V_{10}\over \tilde V_{11}-\tilde V_{00}}f =\eps f.
\end{equation}
Comparing with the zeroth order eigenvalue $\eps_m^{(0)}$
(which does not take into account the mixing), the corrected
eigenvalue for the ground state is then given by
\begin{equation}
\eps_0 \simeq \eps_0^{(0)}+\biggl\langle 
{\tilde V_{01}\tilde V_{10}\over \tilde V_{11}-\tilde
V_{00}}\biggr\rangle
\sim \eps_0^{(0)}+\biggl\langle
{\tilde V_{01}\tilde V_{10}\over
\eps_0^{(0)}-\eps_1^{(0)}}\biggr\rangle,
\end{equation}
where $\langle\cdots\rangle$ denotes expectation value. 
Requiring the second term to be smaller than the first, we have
$\langle\tilde V_{01}\rangle^2/l\ll l^2$, 
where we have used $|\eps_0^{(0)}|\sim l^2$ and
$|\eps_0^{(0)}-\eps_1^{(0)}|\sim l$. Since
\begin{equation}
\langle\tilde V_{01}\rangle
\sim -\biggl\langle{\bf R_{\perp}}\cdot\nabla
{1\over r}\biggr\rangle_{01}
\sim -R_{\perp}\biggl\langle{x\over r^3}\biggr\rangle_{01}
\sim -R_{\perp}\hat\rho\biggl\langle{1\over r^3}\biggr\rangle
\sim -{1\over L_z\hat\rho}R_{\perp},
\end{equation}
the condition for the ansatz to be valid is 
$R_{\perp}\ll l^{1/2}\hat\rho\,$, i.e., 
the proton transverse displacement must be smaller than 
$\sim\hat\rho$.

\subsubsection{Ansatz B}

At large $R_{\perp}$, the molecule should become two
individual atoms (or atom plus ion). Here we set up a
rectangular coordinate system
so that the two protons are located at $(0,0,Z/2)$ and
$(R_{\perp},0,-Z/2)$. 
The electron wavefunction of H$_2^+$ is assumed to be 
$\Phi_{m0}({\bf r})=W_m(\brp)f_{m0}(z)$, i.e.,
the electron cloud is centered on one of the protons.
Then the problem is essentially equivalent to
calculating how an atom is affected by an external ion. 
The Schr\"odinger equation (3.4) still applies except that the
potential $\tilde V_m(z)$ is replaced by
\begin{equation}\tilde V_{mm}'(z,R_{\perp})=V_m\biggl(|z-{Z\over
2}|\biggr)
+V_{mm}\biggl(|z+{Z\over 2}|,R_{\perp}\biggr),\end{equation} 
where the function $V_m$ and $V_{mm}$ are defined in Eqs.~(3.6) and
(3.24) respectively. The eigenvalue can again be solved and thus the
total energy $E(Z,R_{\perp})$ can be obtained. 

In this ansatz, since the electron wavefunction is
not symmetric with respect to $z\rightarrow -z$, the numerical 
method used in Sec. III.A and III.B.1 (see Paper I) need 
modification. Here we integrate the equation from 
both $\infty$ and $-\infty$. The eigenvalue is obtained by 
matching the solution at $z=0$ (see \cite{Press87}). 
We also note that the classical quadrupole
formula for the ion-atom interaction is not applicable here, 
since we always consider $R_{\perp}\lo L_z$ for a bound state. 

For H$_2$, we choose the two electron orbitals centered on each of the 
protons:
\begin{eqnarray}
\Phi_{m_1}(\br) &=&W_{m_1}(\brp)f_{m_10}(z),\\
\Phi_{m_2}(\br) &=&W_{m_2}(\brp-{\bf R_{\perp}})
f_{m_20}(z) e^{-i BR_{\perp}y/2}.
\end{eqnarray}
The extra factor $e^{-i BR_{\perp}y/2}$
in $\Phi_{m_2}(\br)$ comes from a gauge transformation, so that 
the displaced Landau wavefunction 
$W_{m_2}(\brp-{\bf R_{\perp}})$ is still an eigenstate of
the magnetic Hamiltonian with a fixed gauge (Eq.~[3.2]), i.e.,
\begin{equation}
H_B [W_{m_2}(\brp-{\bf R_{\perp}})e^{-i BR_{\perp}y/2} 
\chi(\downarrow)]=0.
\end{equation}
With this ansatz for the basis wavefunctions, 
the HF equations given in Sec.~III.A (Eq.~[3.12]) can be 
applied, except that $\tilde V_m(z)$ must be replaced by 
$\tilde V_{mm}'(z,R_{\perp})$ given in Eq.~(3.32). 
Also, the direct and exchange kernels (Eqs.~[3.15]-[3.16]) are
replaced by 
\begin{eqnarray}
\tilde D_{m_1m_2}(z_1-z_2,R_{\perp}) 
&=& \int\! d^2{\bf r_{1\perp}}d^2{\bf r_{2 \perp}}
|W_{m_1}({\bf r_{1\perp}})|^2|W_{m_2}({\bf r_{2\perp}}-{\bf
R_{\perp}})|^2
{1\over r_{12}},\\
\tilde E_{m_1m_2}(z_1-z_2,R_{\perp}) 
&=& \int\! d^2{\bf r_{1\perp}}d^2{\bf r_{2\perp}}
W_{m_1}({\bf r_{1\perp}})W_{m_2}({\bf r_{2\perp}}-{\bf R_{\perp}})
W_{m_1}^*({\bf r_{2\perp}})W_{m_2}^*({\bf r_{1\perp}}-{\bf R_{\perp}})
\nonumber\\
& &\times e^{i BR_{\perp}(y_1-y_2)/2}{1\over r_{12}}.
\end{eqnarray}
The function $\tilde D_{m_1m_2}(z,R_{\perp})$ can be expressed
as a sum of the function $V_{mm}$ (see Appendix A)
\begin{equation}
\tilde D_{m_1m_2}(z,R_{\perp})=
\sum_{s=0}^{m_1+m_2}d_s(m_1,m_2){1\over\sqrt{2}}
V_{ss}\biggl({z\over\sqrt{2}},{R_{\perp}\over\sqrt{2}}\biggr),
\end{equation}
thus it can be evaluated using Eq.~(3.24). 
For $R_{\perp}>>\hat\rho$, the exchange interaction between electrons
can be neglected since the electron clouds are separated, 
i.e., we can set $\tilde E_{m_1m_2}(z,R_{\perp})=0$.
Therefore, we only need to solve the corresponding Hartree equations:
\begin{equation}
\biggl[-{\hbar^2\over 2m_e\hat\rho^2}{d^2\over
dz^2}-{e^2\over\hat\rho}
\tilde V_{mm}'(z,R_{\perp})+{e^2\over\hat\rho}\tilde K_m(z,R_{\perp})
-\eps_m\biggr]f_{m0}(z)=0,~~~m=m_1,~m_2,
\end{equation}
where $\tilde K_m$ is given by
\begin{equation}
\tilde K_{m_1}(z,R_{\perp})=\int\!dz'f_{m_20}(z')^2 \tilde D_{m_1m_2}
(z-z',R_{\perp}),
\end{equation}
and similarly for $\tilde K_{m_2}$. 

In Fig.~3, we show the energy curve for H$_2^+$ at $B_{12}=1$. The
electron is assumed to be in the $m=0$ state. The electronic energy
curves $E(Z,R_o)$ are calculated using ansatz A with a fixed value of
$R_\perp=R_0$. Each curve has a minimum at $Z=Z_{eq}(R_o)$. We see
that this equilibrium position is almost independent of $R_o$, i.e., 
$Z_{eq}(R_o)\simeq Z_{eq}(0)=Z_o$.
The curves $E(Z_o,R_\perp)$ with a fixed value of $Z_o$
are calculated using both ansatzs discussed above.
For $R_\perp$ less than a few times $\hat\rho$, ansatz A yields 
lower energy, while for larger $R_\perp$, ansatz B gives the 
correct behavior for the energy curve, i.e., $E(Z_o,R_\perp)
\rightarrow
E_a$ as $R_\perp$ increases. Similar behavior for H$_2$ can also be 
obtained. It is evident from Fig.~2 that 
the curves $E(Z_o,R_\perp)$ are much steeper than $E(Z,R_o)$.
Thus the molecule is tied much more ``rigidly'' to the magnetic field
line than along the field axis.

\section{\bf RESULTS FOR THE MOLECULAR EXCITATION LEVELS}

In this section, we present our numerical results for the excitation 
levels of H$_2$. The results for H$_2^+$ are also included 
for completeness and for comparing with previous calculations
(no previous results for H$_2$ are available). 

\subsection{\bf Electronic Excitations}

The equilibrium electronic state is determined by the minimum 
in the energy curve $E(Z,0)$ (cf.~Fig.~1). 
For H$_2^+$, the electronic state is characterized by a single
quantum number $m$.  For H$_2$, there are two types of 
electronic excitations: the ``tightly-bound'' levels correspond to  
electrons in the $(m,\nu)=(m_1,0)$ and $(m_2,0)$ orbitals, 
and the ``weakly-bound'' excitation corresponds to
$(m,\nu)=(0,0),~(0,1)$. 
We have calculated all the electronic bound states
of H$_2^+$ and H$_2$ for $0.1\le B_{12}\le 10$. 
The results for H$_2$ are summarized
in Table II (for the tightly-bound states) and 
Table III (for the weakly-bound state), while the results
for H$_2^+$ are given in Table IV.
Here, by ``bound'' we mean that the equilibrium electronic energy 
$E_m$ of the molecule is less than $E_a\equiv E_a(0)$, 
the energy of a single atom in the ground state 
(for H$_2^+$), or $2E_a$, the energy of two atoms (for H$_2$). 
Clearly, H$_2$ has more electronic excitation levels than H$_2^+$.
As $B$ increases, the number of bound levels in H$_2$ increases. 
For $B_{12} \le 10$, only single-excitation tightly-bound levels,
i.e., those with $m_1=0$, are bound. The double-excitation levels,
such as $(1,2)$ are not bound until the field strength increases to
$B_{12}\go 50$. Excluding the zero-point oscillation energy of the
protons (see Sec.~V), the dissociation energy the H$_2$ molecule
is given by $D^{(\infty)}=2E_a-E_m$. 

We have also calculated the ground-state energy of the molecule 
in the stronger field regime. For $B_{12}\go 10$, our numerical
results can be well fitted to the form:
\begin{equation}
E_m\simeq -0.091\,(\ln b)^{2.7}~({\rm a.u.}).
\end{equation}
More general fitting formula for $D^{(\infty)}$ is given in 
Eq.~(5.2). 

We note that as $B$ increases, the energy $|E_m|$ of the
tightly-bound levels of H$_2$ increases rapidly, while that of
the weakly-bound level does not change appreciably. 
For $l=\ln b>>1$, the weakly-bound state is indeed an excited state
of the H$_2$ molecule. 
For $B_{12}\lo 0.2$, however, we find that the weakly-bound state
actually has lower energy than the tightly-bound level
$(m_1,m_2)=(0,1)$. Thus for such relatively small magnetic field
strength, the weakly-bound state is the actual ground state of the
molecule. 

\subsection{\bf Aligned Vibrational Excitations}

In the standard Born-Oppenheimer approximation, the Hamiltonian 
describing the relative motion of the protons is simply
$H_i={\bf P}^2/(2\mu)+U(Z,R_{\perp})$,
where $\mu$ is the reduced mass of the proton-pair, and
the interatomic interaction potential $U$ is given by the total
electronic energy $E(Z,R_{\perp})$, as calculated in Sec.~III.
For the vibrations along the $z$-direction, there is no magnetic
force on the protons, and Eq.~(4.1) is a good approximation. 
The aligned vibrations are governed by the potential $U(Z,0)=E(Z,0)$, 
which we can fit to a Morse potential (e.g., \cite{Landau77})
\begin{equation}
U(Z,0)=D_m\left\{1-\exp [-\beta (Z-Z_o)]\right\}^2+E_m,
\end{equation}
where $\beta$ is a fitting parameter, and
\begin{equation}
D_m\equiv U(\infty,0)-E_m.
\end{equation}
Thus $D_m=E_a(m)-E_m$ for H$_2^+$, and $D_m=E_a(m_1)+E_a(m_2)-E_m$ for
H$_2$ (we consider the tightly-bound states only).
The aligned vibrational energy levels are then given by
\begin{equation}
E_{n_{\parallel}}=\hbar\omega_{\parallel}\biggl(n_{\parallel}
+{1\over 2}\biggr)
-{(\hbar\omega_{\parallel})^2\over 4D_m}\biggl(n_{\parallel}
+{1\over 2}\biggr)^2,
\end{equation}
where the vibrational energy quanta is 
\begin{equation}
\hbar\omega_{\parallel}=\hbar \beta\biggl({2D_m\over\mu}\biggr)^{1/2}.
\end{equation}

The values of $\hbar\omega_{\parallel}$ and $D_m$ 
for different bound electronic states and different magnetic field
strength are given in Table II for H$_2$ and in Table IV for H$_2^+$.
In Fig.~1, the numerical $E(Z,0)$ curve is compared with the 
fitted curve (Eq.~[4.2]) for the $(0,1)$ state of H$_2$ at $B_{12}=1$. 
We see that the fitting is indeed very good, especially for the bound
region (below the dark line in Fig.~1). For H$_2$, our results for 
$\hbar\omega_{\parallel}$ are accurate to within about $5\%$. 
The Morse potential fits the $E(Z,0)$ curves of H$_2^+$ less well,
but the resulting $\hbar\omega_{\parallel}$ is still accurate to within
about $10\%$. 

As discussed in Sec.~II, $\hbar\omega_{\parallel}$ (cf.~Eq.~[2.8])
is approximately proportional to $(\ln b)^2\mu^{-1/2}$
times a slowly increasing function of $B$. Our numerical results 
confirm this approximate scaling relation. A better empirical
scaling is $\hbar\omega_{\parallel} \propto (\ln b)^{5/2}\mu^{-1/2}$.
Thus for the $(m_1,m_2)=(0,1)$ state of H$_2$ ($\mu\simeq 918$), 
we have
\begin{equation}
\hbar\omega_{\parallel} \simeq 0.13 (\ln b)^{5/2}\mu^{-1/2} ({\rm a.u.})
\simeq 0.12 (\ln b)^{5/2}~({\rm eV}),~~~~~({\rm H}_2).
\end{equation}
For the ground state ($m=0$) of H$_2^+$, we have
\begin{equation}
\hbar\omega_{\parallel} \simeq 0.085 (\ln b)^{5/2}\mu^{-1/2} 
({\rm a.u.})
\simeq 0.076 (\ln b)^{5/2}~({\rm eV}),~~~~~
({\rm H}_2^+).
\end{equation}
Both Eqs.~(4.6) and (4.7) are accurate to within about $10\%$. 
These fitting expressions are indeed very satisfactory considering
the approximation introduced when we use Morse potential to
fit the numerical $E(Z,0)$ curves.

There is no previous reliable calculation for H$_2$ molecules. 
For H$_2^+$, our results for the ground state electronic energy, 
interatomic spacing and aligned vibrational energy quanta 
$\hbar\omega_{\parallel}$ agree with those obtained by 
Wunner, Herold and Ruder \cite{Wunner82}, and those of
Le Guillou and Zinn-Justin \cite{LeGuillou84}, who used a
similar method as ours in the aligned cases. The slight difference
in $\hbar\omega_{\parallel}$ between our results and theirs 
is likely due to the different ways
of extracting this quantity: we obtain it by fitting 
$E(Z,0)$ to a Morse potential, while they obtained it by  
evaluating the second derivative of $E(Z,0)$ 
around the equilibrium separation. 
Le Guillou and Zinn-Justin also considered the effects of 
non-adiabaticity (i.e., mixing of different electron Landau levels).
This is negligible for field strength of interest in this paper 
($b\gg 1$). The variational calculation of Khersonskii
\cite{Khersonskii84}
gave somewhat smaller (by about $20\%$) values for
$\hbar\omega_{\parallel}$.
This is due to the inaccuracy in his atomic binding energy.

\subsection{\bf Transverse Vibrational Excitations}

Neglecting the magnetic forces on the protons, the transverse
oscillations of the molecule are governed by the potential 
$U(Z_o,R_{\perp})=E(Z_o,R_{\perp})$. Our calculation of this function
is less accurate than the aligned case, and yields only an
{\it upper limit} to the exact potential. For small-amplitude
oscillation (see Sec.~II), we fit this potential to a harmonic form
\begin{equation}
\delta U(R_{\perp})=U(Z_o,R_{\perp})-U(Z_o,0)
\simeq {1\over 2}\mu\,\omega_{\perp 0}^2 R_{\perp}^2.
\end{equation}
The transverse vibrational motion of the protons is therefore
described by a two-dimensional harmonic oscillator. The numerical
values for the transverse vibrational energy quanta are tabulated
in Table II (for H$_2$) and in Table IV (for H$_2^+$). Only the
results for the ground electronic states are given. 

Note that $\hbar\omega_{\perp 0}$ is larger than 
$\hbar\omega_{\parallel}$ for $B_{12}\go 1$. Also, 
$\hbar\omega_{\perp 0}$ and $\hbar\omega_{\parallel}$ can be
comparable or even larger than the
electronic excitation energy spacings $\Delta E_m$. 
This is in contrast with the zero-field cases, 
where one has $\Delta E_m>>\hbar\omega_{vib}
>>\hbar\omega_{rot}$. Although the actual values of
$\hbar\omega_{\perp 0}$ may be somewhat smaller than our results, the
qualitative features revealed in our calculations are expected to be
valid in general. 

The discussion in Sec.~II gives 
$\hbar\omega_{\perp 0}\sim (\xi^{-1}b\ln b)^{1/2}\mu^{-1/2}$
(Eq.~[2.10]), where $\xi^{-1}$ increases slowly with increasing $B$. 
Our numerical results confirm this approximate scaling relation
and $\hbar\omega_{\perp 0}\propto b^{1/2}(\ln b)\mu^{-1/2}$
fits better the results in Table II and Table IV.
For the $(m_1,m_2)=(0,1)$ state of H$_2$, we have
\begin{equation}
\hbar\omega_{\perp 0}\simeq 0.125 b^{1/2}(\ln b)\mu^{-1/2} 
({\rm a.u.})
=0.553\left({b\over b_{crit}}\right)^{1/2}\!\!\ln b\,({\rm a.u.})
=0.11~b^{1/2}(\ln b)~({\rm eV}),
~~~~({\rm H}_2).
\end{equation}
For the $m=0$ state of H$_2^+$, we have
\begin{equation}
\hbar\omega_{\perp 0}\simeq 0.14 b^{1/2}(\ln b)\mu^{-1/2} ({\rm a.u.})
=0.62\left({b\over b_{crit}}\right)^{1/2}\!\!\ln b\,({\rm a.u.})
=0.13~b^{1/2}(\ln b)~({\rm eV}),
~~~~({\rm H}_2^+).
\end{equation}

Our results for $\hbar\omega_{\perp 0}$ of H$_2^+$ also agree closely
with those of Le Guillou and Zinn-Justin \cite{LeGuillou84} obtained
using their ``static approximation'', which is similar to ansatz A
adopted in our paper (Sec. III.B.1). Their improved calculations
indicate that the real value of $\hbar\omega_{\perp 0}$ can be lower
by tens of percent (from about $10\%$ for $B_{12}=0.1$ to about
$40\%$ for $B_{12}=5$). 
We expect our results for H$_2$ to have similar accuracy. 
However, as noted in Sec.~II.C, the present results apply only to the 
small-amplitude ($R_{\perp}\lo\hat\rho$) vibrations and relatively
weak field ($B\lo B_{crit}$).
For sufficiently large field strength, the magnetic forces on 
the protons become important and can change the transverse vibration 
energy significantly, as we discuss below.

\section{\bf EFFECTS OF FINITE PROTON MASS ON THE ELECTRONIC ENERGY
AND MOLECULAR DISSOCIATION ENERGY}

Our calculations and results in the previous sections are based on 
Born-Oppenheimer approximation where the proton positions
are fixed when we consider the electronic energy of the molecule.
For finite proton mass, one can rescale the electronic energy
by replacing the electron mass with an appropriate reduced mass.
This only introduces a small correction (of order $m_e/m_p$),
and is neglected in our paper. However, as noted in Sec.~I,
the separation of the proton and electron motion 
in strong magnetic field is much more complicated, especially
in the regime of $B\go B_{crit}$ when the cyclotron energy
of proton,
\begin{equation}
\hbar\omega_p=\hbar(eB/m_pc)=(b/b_{crit})\ln b_{crit}\,({\rm a.u.})
=6.3 B_{12}\,({\rm eV}),
\end{equation}
is comparable with or larger than the spacing of the electronic energy
levels. The ``standard procedure'' for separating the proton and
electron motion leads to some ambiguities regarding
the binding energy of H$_2$ in the strong field regime; these are 
discussed in Sec.~V.A. An alternative scheme, which is more suitable
for $B\go B_{crit}$, is described in Sec.~V.B. An approximate
expression for the ``corrected'' dissociation energy of H$_2$ in the
ground state is given by Eq.~(5.7). 

\subsection{\bf Unbound States from the Standard Scheme When 
$B\go B_{crit}$}

In Sec.~II and III we have followed the ``standard procedure'' for 
molecules, where one first considers the two protons as infinitely
massive fixed at equilibrium separation $Z_o$ along 
the same field line, with their motion included only as an 
``afterthought''. This is strictly valid only for $B << B_{crit}$,
where the zero-point vibration amplitudes and the magnetic
force on the protons are small. The two electrons, 
both in the lowest Landau level, are in cylindrical orbitals $m_1$
and $m_2$ centered on the proton field line
with radii given by equation (1.2). The Pauli principle
requires $m_1\ne m_2$ so that the ground state has $m_1+m_2=1$. As the
p-p separation $Z$ is allowed to increase, the system tends 
to two free H atoms, one in orbital state $m_1$ and another $m_2$.
The standard procedure for treating the two-body
problem of hydrogen atom\cite{Lai95a,Avron78} deals with states 
where the transverse pseudomomentum of each atom is zero, in which
case the protons must have Landau excitations $m_1$ and $m_2$,
respectively. The simplest state for the molecule with
electronic orbitals $m_1$ and $m_2$ is then the state where the
protons have these Landau excitations at all separations (even though
the transverse pseudomomentum is conserved only for the total
molecule, not individual atoms). This choice adds
a Landau excitation energy $(m_1+m_2)\hbar\omega_p$,
i.e., even the electronic ground state would have an additional
positive energy $\hbar\omega_p$ and would be unbound
(relative to two ground-state atoms) for $B>>B_{crit}$. This
molecular state has the simplest wavefunction but not necessarily 
the lowest energy, since there are states where pseudomomentum is not
zero (corresponding to finite separation of the guiding centers of 
the electron and proton; see Ref.\cite{Lai95a}). States without the
additional energy $\hbar\omega_p$ are discussed in Sec.~V.B. 

For infinite proton mass, the dissociation energy of H$_2$
is given by $D^{(\infty)}=2E_a(0)-E_m$. Our numerical results for 
the $(m_1,m_2)=(0,1)$ ground state can be written in the following
form
\begin{equation}
D^{(\infty)}\simeq 0.106\left[1+\tau\ln\left({b\over b_{crit}}
\right)\right](\ln b)^2~({\rm a.u.}),~~~~
\tau\simeq 0.1\,(\ln b)^{0.2},
\end{equation}
where $\tau$ varies slowly with $b$ ($\tau\simeq 0.14$ for 
$b\sim 10^3$ and $\tau\simeq 0.17$ for $b\sim 10^5$).
For field strength $b\simeq b_{crit}$
the square bracket in Eq.~(5.2) can be replaced by unity. 
As shown in Sec.~II.B and Sec.~IV.B, the aligned proton vibrations
have an energy spacing of order
$\hbar\omega_\parallel\sim\mu^{-1/2}D^{(\infty)}$
and a small vibration amplitude of order $\delta Z\sim \mu^{-1/4}Z_o$, 
where $Z_o$ is the equilibrium separation between the protons 
(Eq.~[2.5]). The inequality $\hbar\omega_\parallel<<D^{(\infty)}$ 
does not depend appreciably on the 
magnetic field strength, so for the ground 
molecular state we should be able to use $\hbar\omega_\parallel$ 
in Eq.~(4.6) for the aligned vibrations
even when $B>>B_{crit}$; furthermore, since $\delta Z<<Z_o$, we 
do not need to consider the Pauli principle explicitly for the
transverse wavefunctions of the protons. 
For treating this transverse motion, however, the magnetic force
becomes important when $b\go b_{crit}$, as can be seen from the 
ratio of the expressions in Eq.~(5.1) and (4.9), 
$\hbar\omega_p/\hbar\omega_{\perp 0}\simeq
1.81(b/b_{crit})^{1/2}(\ln b_{crit}/\ln b)$. 
We can give at least a plausibility argument for the inclusion 
into the ``standard scheme'' of the
magnetic effects on the transverse motion from the following
consideration: A free proton in the magnetic field $B$ has a 
zero-point energy $\hbar\omega_p/2$. 
This can be thought of as the ground state energy in a ``magnetic 
restoring potential'' $(1/2)m_p\omega_p^2(R_\perp/2)^2$, 
which gives a ground-state wavefunction of size 
$R_\perp\sim\hat\rho$ (independent of mass)
as in the Landau wavefunction (Eq.~[2.2]). Thus the total restoring 
potential for for the protons in H$_2$ is given by the sum of 
the ``electronic potential'' $\delta U(R_\perp)$, which we have
calculated in Sec.~III, and twice (for two protons) of the magnetic
restoring potential. For $R_\perp\lo\hat\rho$, we have seen that 
$\delta U(R_\perp)$ can be approximated
by the quadratic form in Eq.~(4.8), so that the total potential 
is $\mu (\omega_{\perp 0}^2+\omega_p^2)R_\perp^2/2$. The size of the 
ground-state wavefunction is then $\lo \hat\rho$, the approximation is
justified and the excitation energy quanta is 
$\hbar (\omega_{\perp 0}^2+\omega_p^2)^{1/2}$. 
Since the energy of $\hbar\omega_p/2$ also exists in isolated H atom, 
the zero-point energy for the transverse oscillation of the molecule 
can be written as 
\begin{equation}
\hbar\omega_\perp=\hbar (\omega_{\perp 0}^2+\omega_p^2)^{1/2}
-\hbar\omega_p.
\end{equation}
For $b<<b_{crit}$ we have $\hbar\omega_\perp\simeq
\hbar\omega_{\perp 0}$ as expected, and Eq.~(5.2) shows
that $\hbar\omega_\perp/D^{(\infty)}\sim 5.(b/b_{crit})^{1/2}
/\ln b<<1$. For $b>>b_{crit}$, on the other hand, 
$\hbar\omega_\perp\simeq \hbar\omega_{\perp 0}^2/(2\omega_p)
\simeq 0.016\,l^2$ so that $\hbar\omega_\perp/D^{(\infty)}
\sim 0.15$. Thus the transverse zero-point energy $\hbar\omega_\perp$
remains less than the dissociation energy for the state
given by the standard scheme, but the Landau energy $\hbar\omega_p$
of the excited proton has to be added to the molecular energy
also. The ``corrected'' dissociation energy in this 
scheme is then given by
\begin{equation}
D^{(std)}=D^{(\infty)}-{1\over 2}\hbar\omega_\parallel-
\Delta\eps^{(std)},~~~~~
\Delta\eps^{(std)}=\hbar (\omega_{\perp 0}^2+\omega_p^2)^{1/2}.
\end{equation}
Clearly, $D^{(std)}$ becomes negative (i.e., the state is unbound)
as $b$ increases beyond $b_{crit}$. 
We shall see in Sec.~V.B that an alternative scheme gives molecular 
{\it bound} states with lower energy for $b\go b_{crit}$. 

\subsection{\bf The Alternative Scheme}

The alternative scheme we propose for the H$_2$ molecule ground state
is a generalization of the scheme for H atom described in Sec.~IV of
Ref.\cite{Lai95a}. In this scheme the transverse pseudo-momentum is
not chosen as a good quantum number, and our approximate wavefunction
will not be an exact eigenstate of the Hamiltonian. However, 
it does provide a suitable trial wavefunction 
and enable us to obtain a rigorous {\it lower}
limit to the dissociation energy $D$. 
In the absence of Coulomb interaction between the four 
particles, there are eight quantum
numbers specifying the transverse degrees of freedom of the system: 
the Landau excitation number $n$ and the orbital number $m$ for each 
of the four particles. For the H$_2$ ground state we then choose 
$n=0$ for both electrons and both protons (so there is no
$\hbar\omega_p$ contribution to the electronic energy of the
molecule), and $(m_1,m_2)=(0,1)$ for the electrons. We can 
choose $m=0$ for both protons since, as mentioned, 
the proton $z$-wavefunction can be anti-symmetrized to satisfy 
the Pauli principle with little energy contribution. 
As a trial wavefunction we assume that 
the charge distribution of protons consists of two 
sheets separated in the $z$-axis by distance $Z$, with 
surface density given by $|W_0(\br_{\perp})|^2$. Obviously, when 
the Coulomb potentials between the particles are restored, $n$ and $m$
for the individual particle cease to be good quantum numbers
\footnote{The only good quantum number for the transverse 
degrees of freedom is the total orbital angular momentum 
along the $z$-axis $L_z=\sum_i {\rm sign}(e_i)(m_i-n_i)$, 
where ${\rm sign}(e_i)=1$ for proton and ${\rm sign}(e_i)=-1$ for
electron\cite{Lai95a}. For $b>>b_{crit}$, the Landau excitation
numbers $n$ for both electrons and protons are ``adiabatically''
conserved and can be set to $0$ for the ground state. In this case 
$L_z=m_{1p}+m_{2p}-m_{1e}-m_{2e}$. Thus the true ground
state of the molecule for $B>>B_{crit}$ involves a mixing of many 
different $(m_{1e},m_{2e},m_{1p},m_{2p})$ states with the same $L_z$.},
but the trial wavefunction thus constructed will give an upper bound
to the true ground-state energy of the molecule according to the
variational principle. This ``trial energy'' can be calculated using
the Hartree-Fock method described in Sec.~II.A, subjected to 
two modifications:
(i) The averaged electron-proton interaction potential $\tilde V_m(z)$
in equation (3.5) is replaced by 
\begin{eqnarray}
\tilde V_m(z) &\rightarrow&\int\!d^2\br_{\perp e}
|W_m(\br_{\perp e})|^2\int\! d^3\br_p|W_0(\br_{\perp p})|^2
\left[\delta\left(z_p-{Z\over 2}\right)+\delta
\left(z_p+{Z\over 2}\right)\right]{1\over |\br_e-\br_p|}\nonumber\\
&=& D_{0m}\left(z-{Z\over 2}\right)+D_{0m}\left(z+{Z\over 2}\right),
\end{eqnarray}
where $D_{0m}(z)$ is defined in Eq.~(3.17);
(ii) The proton-proton interaction term $e^2/Z$ in equation (3.19) 
is replaced by $D_{0m}(Z)$ (although this modification 
has negligible effect on the energy except when $Z\rightarrow 0$). 

The molecular energy $E_m^{(alt)}$ obtained by this alternative scheme
is larger than the result $E_m$ obtained using the scheme of
Sec.~III.A (where the protons are treated as infinitely massive), by
some amount $\Delta\eps^{(alt)}$. The weakening of the electron-proton
interaction is due to the spread of the proton 
wavefunction by an amount of order $\hat\rho$. However, 
since $Z>>\hat\rho$ the change involves only the
logarithm of the Coulomb energy. This can be characterized 
by changing $b$ in Eq.~(4.1) to $b/(2C)$, where $C$ is of order unity,
i.e., $E_m^{(alt)}\simeq -0.091\,[\ln (b/2C)]^{2.7}$.
To leading order in $\ln (2C)$, we then have
\begin{equation}
D^{(alt)}=D^{(\infty)}-{1\over 2}\hbar\omega_\parallel
-\Delta\eps^{(alt)};~~~~~
\Delta\eps^{(alt)}\simeq 0.24\ln(2C)(\ln b)^{1.7}~({\rm a.u.}),
\end{equation}
as an alternative to Eq.~(5.4). 

We have performed numerical calculations and found that
the ``trial'' ground-state energy thus obtained agrees with the 
result using the scheme of Section III.A to within $15\%$.
For $B_{12}=50,100,500,10^3$, we found 
$\Delta\eps^{(alt)}\simeq 162,~191,~258,~294$ eV, corresponding to
$C\simeq 0.8$ for $B_{12}=50$ and $C\simeq 0.9$ for $B_{12}=10^3$. 
The numerical values for $\Delta\eps^{(alt)}$ can be fitted
by $\Delta\eps^{(alt)}\simeq 0.06(\ln b)^2$. This has the same scaling
with $b$ as $\hbar\omega_\perp$ defined in Eq.~(5.3). With this
value of $\Delta\eps^{(alt)}$, the dissociation energy given by 
Eq.~(5.6) is larger than that from Eq.~(5.4), and therefore
represents the true molecular ground state for all $b\go b_{crit}$. 
Our numerical calculation of $\Delta\eps^{(alt)}$ used a particularly
simple trial wavefunction and a better wavefunction with 
the variational method would presumably lower $\Delta\eps^{(alt)}$
somewhat. This would lower the numerical value of $b$ above which the
true ground state is the state with no Landau excitations 
for either proton, obtained by the present ``alternative scheme''. 

The fact that $\Delta\eps^{(alt)}$ scales similarly with $b$ 
as $\hbar\omega_\perp$ suggests that for practical purpose, the 
``corrected'' dissociation energy of H$_2$ in the ground state 
can be approximated by 
\begin{equation}
D\simeq D^{(\infty)}-\left({1\over 2}\hbar\omega_\parallel
+\hbar\omega_\perp\right),
\end{equation}
for all field regimes ($b>>1$), where $\hbar\omega_\perp$ is 
given by Eq.~(5.3). The numerical results for a wide range of field
strength are summarized in Table V.

\section{\bf CONCLUSIONS}

In this paper, we have, for the first time, studied and 
characterized the energy excitation levels of H$_2$ molecule in a 
superstrong magnetic field ($B\go 10^{12}$ G) which exists on the
surfaces of many neutron stars. The main theoretical uncertainty of
our calculations lies in the non-trivial separation of the motion of
the protons and 
that of the electrons. Nevertheless, we find that in such 
a strong magnetic field, H$_2$ molecule exhibits completely different
energy excitation levels as compared to its well-known zero-field
counterpart. The fact that the excitation energies associated with the
oscillations of the protons are comparable to the electronic
excitations indicates
that the statistical weight of a H$_2$ molecule is not much 
larger than that of a H atom. This greatly simplifies 
the calculations of the chemical equilibria of various forms of H in 
a neutron star atmosphere\cite{Lai95b}. 

Larger hydrogen molecules and chains can also form in a superstrong 
magnetic field. Their ground state binding energies have been 
calculated in Paper I. It is expected that these larger molecules 
possess qualitatively similar energy excitation levels as those 
of H$_2$ considered in this paper, with one exception: 
For a long chain molecule H$_n$ with $1<<n<<[b/(\ln b)^2]^{1/5}$,
the spacing $Z_o$ along a field line between adjacent protons
decreases with increasing $n$ approximately as $n^{-2}$.
The fractional zero-point vibration amplitude $\Delta Z/Z_o$ 
is of order $(m_e/m_p)^{1/4}n^{1/2}$. The aligned vibrations thus
become more pronounced as $n$ increases (and can lead to 
``internal pycnonuclear reactions'' which will be discussed
in\cite{Lai95b}). 

There is no question that the exotic molecules considered in this paper
exist on the surfaces of some neutron stars with $B_{12}\go 10^{12}$ G
and temperature $T\sim 10^5-10^6$ K \cite{Lai95b}. For very low
surface temperature
($T\lo 10^5$ K), the atmosphere is likely to condensate into a metallic 
state, since the hydrogen metal has the largest binding energy. 
However, for the astrophysically more interesting temperature range
($T\go 10^5$ K), the outer layer of a neutron star 
will predominantly exists in the form of nondegenerate gas of 
individual atoms and small molecules: e.g., 
when $T\sim 3\times 10^5$ K,
the photosphere of a neutron star is dominated by atoms 
if $B_{12}=1$, while it is dominated by H$_2$ if $B_{12}=10$. 
The existence of H$_2$ in the atmosphere will give rise to 
appreciable radiative opacity. For example, 
since the proton separation in H$_2$ is different from that in H$_2^+$
(see Table II and Table IV), the photo-ionization
cross-section from the ground state of H$_2$ is expected to be small
according to the Franck-Condon principle. However, photo-ionization
from an excited vibrational state or electronic state, for which the 
proton separation is close to that in H$_2^+$ ground state, can provide
significant continuum opacity. These issues may warrant further study,
especially in light of the increasing possibility of
the spectroscopic studies of isolated neutron stars by future X-ray/EUV
satellites.

\acknowledgments
This work has been supported in part by NSF Grants Nos. AST 91--19475
AST 93--15375 and NASA Grant No. NAGW--666 to Cornell University 
and by NASA Grant No. NAGW-2394 to Caltech. 
D.L. also acknowledges the support of 
Richard C. Tolman Research Fellowship in theoretical astrophysics at
Caltech.

\appendix
\section{Coulomb Integrals for Landau Wavefunction}

In this Appendix, we derive Eqs.~(3.24) and (3.38). 
First consider the function
\begin{equation}
V_{mm}(z,r_o)=\langle m|{1\over |\br-\br_o|}|m\rangle,
\end{equation}
with ${\bf r_o}=r_o{\bf \hat x}$. 
Since 
\begin{equation}
{1\over r}={1\over 2\pi^2}\int\! {d^3q\over q^2}e^{i{\bf q}\cdot\br},
\end{equation}
we have 
\begin{equation}
V_{mm}(z,r_o)=
{1\over 2\pi^2}\int\!\!{d^3q\over q^2}e^{-i{\bf q}\cdot{\bf r_o}}
e^{iq_zz}\langle m|e^{i{\bf q_{\perp}}\cdot{\bf r_{\perp}}}|m\rangle.
\end{equation}
Using the general result for the  matrix element \cite{Virtamo75}
\begin{eqnarray}
\langle m'|e^{i{\bf q_{\perp}}\cdot\brp}|m\rangle
&=&(-1)^m i^{m+m'}\biggl({m!\over m'!}\biggr)^{1/2}
e^{-q_{\perp}^2/2}L_m^{m'-m}\biggl({q_{\perp}^2\over 2}\biggr)
\biggl({q_{\perp}\over\sqrt{2}}\biggr)^{m'-m}e^{i(m'-m)\theta_q},
\nonumber\\
& & ~~~~~~~~~~~~~~~~~~~~~~~~~~~(m'\ge m),
\end{eqnarray}
where $\theta_q$ specifies the angle of ${\bf q_{\perp}}$ in the 
$q_x-q_y$ plane, and $L_n^m$ is the Laguerre polynomial of order $n$
\cite{Abramowitz72}, we have
\begin{equation}
\langle m|e^{i{\bf q_{\perp}}\cdot{\bf r_{\perp}}}|m\rangle
=e^{-q_{\perp}^2/2}L_m\biggl({q_{\perp}^2\over 2}\biggr).
\end{equation}
Substitute Eq.~(A5) into (A3), and integrate out $dq_z$ and
$d\theta_q$ using 
\begin{equation}\int\!{dq_z\over q_z^2+q_{\perp}^2}e^{iq_zz}=
{\pi\over q_{\perp}}
e^{-q_{\perp}|z|},
\end{equation}
and 
\begin{equation}
\int\!d\theta_q e^{-iq_{\perp}r_o\cos\theta_q}
=2\pi J_0(q_{\perp}r_o),
\end{equation}
we obtain 
\begin{equation}
V_{mm}(z,r_o)=\int_0^{\infty}\!\!dq_{\perp} 
e^{-q_{\perp}^2/2-q_{\perp}|z|}J_0(q_{\perp}r_o)
L_m\biggl({1\over 2}q_{\perp}^2\biggr),
\end{equation}
i.e., Eq.~(3.24). Note that using Eq.~(A4), a more general expression
can be obtained for the matrix element
\begin{eqnarray}
&&V_{m'm}(z,r_o) \equiv \langle m'|{1\over |\br-\br_o|}|m\rangle
\nonumber\\
&&~~~~=\biggl({m!\over m'!}\biggr)^{1/2}\int_0^{\infty}\!\!dq
\biggl({q\over\sqrt{2}}\biggr)^{m'-m}
e^{-q^2/2-q|z|}J_{m'-m}(qr_o)L_m^{m'-m}\biggl({1\over 2}q^2\biggr),
~~~~~(m'\ge m).
\end{eqnarray}

Next consider $\tilde D_{m_1m_2}$ defined in Eq.~(3.36). Changing
variable
$({\bf r_{2\perp}}-{\bf R_{\perp}})\rightarrow {\bf r_{2\perp}}$, we
have
\begin{equation}
\tilde D_{m_1m_2}(z_1-z_2,R_{\perp}) 
=\int\! d^2{\bf r_{1\perp}}d^2{\bf r_{2 \perp}}
|W_{m_1}({\bf r_{1\perp}})|^2|W_{m_2}({\bf r_{2\perp}})|^2
{1\over |\br_1-\br_2-{\bf R_{\perp}}|}.
\end{equation}
Using Eq.~(A2), we have
\begin{equation}
\tilde D_{m_1m_2}(z,R_{\perp})
={1\over 2\pi^2}\int\!\!{d^3q\over q^2}e^{-i{\bf q}\cdot
{\bf R_{\perp}}}
e^{iq_zz}\langle m_1|e^{i{\bf q_{\perp}}\cdot
{\bf r_{\perp}}}|m_1\rangle
\langle m_2|e^{i{\bf q_{\perp}}\cdot{\bf r_{\perp}}}|m_2\rangle.
\end{equation}
Again, using Eq.~(A4), and integrating out $dq_z$ and $d\theta_q$ with
Eqs.~(A6)-(A7), we obtain
\begin{equation}
\tilde D_{m_1m_2}(z,R_{\perp})
=\int_0^{\infty}\!\!dq e^{-q^2-q|z|}J_0(qR_{\perp})
L_{m_1}\biggl({1\over 2}q^2\biggr) L_{m_2}\biggl({1\over 2}q^2\biggr).
\end{equation}
Now defining the coefficient $d_s(m_1,m_2)$ via (see Paper I)
\begin{equation}
L_{m_1}({x \over 2})L_{m_2}({x \over 2})=\sum_{s=0}^{m_1+m_2}
d_s(m_1,m_2) L_s(x),
\end{equation}
Eq.~(A12) then becomes
\begin{eqnarray}
\tilde D_{m_1m_2}(z,R_{\perp})
&=&\sum_{s=0}^{m_1+m_2}d_s(m_1,m_2)
\int_0^{\infty}\!\!dq e^{-q^2-q|z|}J_0(qR_{\perp})L_s(q^2)\nonumber\\
&=&\sum_{s=0}^{m_1+m_2}d_s(m_1,m_2) 
\int_0^{\infty}\!\!{dq\over\sqrt{2}}e^{-q^2/2-q|z|/\sqrt{2}}
J_0\biggl({qR_{\perp}\over\sqrt{2}}\biggr)L_s\biggl({q^2\over 2}\biggr),
\end{eqnarray}
which reduces to Eq.~(3.38) after using Eq.~(A8). 


\section{More Accurate Calculation of H$_2^+$}

An ``exact'' treatment of H$_2^+$ for general orientation 
of the molecular axis proceed as follows. Consider 
the coordinate system of Ansatz A in Sec.~III.B.1. 
When $b>>1$, the most general electron wavefunction for the $\nu=0$
state can be written as
\begin{equation}
\Phi_{m0}(\br)=\sum_m W_m(\brp)f_{m0}(z).
\end{equation}
Substituting this into the Schr\"odinger equation and averaging over
the transverse direction, we obtain a set of differential 
equations for $f_{m0}(z)$:
\begin{equation}
-{\hbar^2 \over 2m_e\hat\rho^2}{d^2\over
dz^2}f_{m0}(z)-{e^2\over\hat\rho}
\sum_{m'}\tilde V_{mm'}(z,R_{\perp/2})f_{m'0}(z)=\eps_{m0} f_{m0}(z),
~~~~~~m=0,1,\cdots
\end{equation}
where $V_{mm'}$ is defined similar to Eq.~(3.23):
\begin{equation}
\tilde V_{mm'}(z,R_{\perp}/2)=
V_{mm'}\biggl(|z-{Z\over 2}|,{R_{\perp}\over 2}\biggr)
+V_{mm'}\biggl(|z+{Z\over 2}|,{R_{\perp}\over 2}\biggr),
\end{equation}
and the function $V_{mm'}$ can be evaluated using Eq.~(A9). 
Equation (B2) is subject to the boundary conditions 
$df_{m0}/dz=0$ at $z=0$ and
$f_{m0}\rightarrow 0$ as $z\rightarrow\infty$. The normalization 
condition requires
\begin{equation}
\sum_m\int_{-\infty}^{\infty}\! dz |f_{m0}(z)|^2=1.
\end{equation}

The set of equations (B2) can be solved numerically using an iterative
scheme similar to that used for solving the Hartree-Fock equation
(Paper I). Successively accurate results can be obtained by using
increasing number of
terms in the sum in Eq.~(B1). The lowest energy state corresponds to 
the solution satisfying
\begin{equation}
\int_{-\infty}^{\infty}\! dz |f_{00}(z)|^2
>\int_{-\infty}^{\infty}\! dz |f_{10}(z)|^2
>\int_{-\infty}^{\infty}\! dz |f_{20}(z)|^2
>\cdots.
\end{equation}
Generalization of this method to H$_2$ molecule is much more
complicated. 


\bigskip

\begin{figure}
\caption
{The electronic energy curves $E(Z,0)$ for the tightly-bound states 
of H$_2$ molecule at $B=10^{12}~{\rm G}$ 
when the molecular axis is aligned with the magnetic field axis.
The electrons occupy the $(m,\nu)=(m_1,0)$ and $(m_2,0)$ orbitals
($m_1\neq m_2$). The solid line is for the state $(m_1,m_2)=(0,1)$,
the short-dashed line for $(0,2)$, the long-dashed line for $(0,3)$. 
The dotted line is from the fitting using the Morse potential
(Eq.~[4.2]). 
The dark solid line corresponds to the energy of two isolated H atoms 
in the ground state $2E_a=-323$ eV.
}
\end{figure}

\begin{figure}
\caption
{The electronic energy curves $E(Z,0)$ of H$_2$ molecule 
at $B=10^{12}~{\rm G}$ 
when the molecular axis is aligned with the magnetic field axis.
The solid line corresponds to the tightly-bound state in which the
electrons 
occupy the $(m,\nu)=(0,0)$ and $(1,0)$ orbitals, the dashed line
corresponds to the weakly-bound state in which the electrons occupy
the $(0,0)$ and $(0,1)$ orbitals. 
}
\end{figure}

\begin{figure}
\caption{
The electronic energy curves for the ground state of H$_2^+$ 
at $B=10^{12}~{\rm G}$. 
The light lines show the $E(Z,R_o)$ curves
with a fixed $R_\perp=R_o$ for $R_o=0$ (solid line),
$R_o=\hat\rho$ (dotted line) and $R_o=2\hat\rho$ (dashed line). 
The dark lines show the function $E(Z_o,R_\perp)$ for a fixed value of
$Z_o$ given by the equilibrium separation of the protons. 
The solid line is calculated using ansatz A, the dotted line using
ansatz B (see Sec.~III.B). 
}
\end{figure}

\newpage

\mediumtext
\begin{table}
\caption{Energy levels $E_a(m)$ (in eV) of hydrogen atom in 
superstrong magnetic field.
The levels are specified by the quantum number $m$, while the 
longitudinal node $\nu=0$. Here $B_{12}=B/(10^{12}~{\rm G})$.}
\begin{tabular}{c c c c c c c}
$B_{12}$ &$E_a(0)$ &$E_a(1)$ &$E_a(2)$ &$E_a(3)$ &$E_a(4)$ & $E_a(5)$\\
\hline
$0.1$&$-76.4$&$-52.5$&$-43.3$&$-38.0$&$-34.4$&$-31.8$\\
$0.5$&$-130.2$&$-92.8$&$-77.8$&$-69.0$&$-63.0$&$-58.6$\\
$1$&$-161.5$&$-116.9$&$-98.7$&$-88.0$&$-80.6$&$-75.1$\\
$2$&$-198.5$&$-145.8$&$-124.1$&$-111.2$&$-102.2$&$-95.5$\\
$5$&$-257.1$&$-192.6$&$-165.5$&$-149.2$&$-137.8$&$-129.2$\\
$10$&$-309.6$&$-235.1$&$-203.5$&$-184.3$&$-170.9$&$-160.7$\\
\end{tabular}
\end{table}

\mediumtext
\begin{table}
\caption{The tightly-bound energy levels of H$_2$ molecule in which 
the electrons occupy the $(m,\nu)=(m_1,0)$ and $(m_2,0)$ orbitals
($m_1\neq m_2$). Here $B_{12}=B/(10^{12}~{\rm G})$,
$E_a$ is the ground state energy of H atom, 
$m_1,m_2$ are the quantum numbers specifying the electronic
excitations,
$E_m$ is the electronic energy of the molecule,
$Z_o$ is the equilibrium interatomic separation ($a_o$ is the Bohr
radius),  
$D_m$ is defined by $D_m\equiv E_a(m_1)+E_a(m_2)-E_m(m_1,m_2)$, 
$\hbar\omega_{\parallel}$ is the aligned vibrational energy quanta,
and $\hbar \omega_{\perp 0}$ is the transverse vibrational energy
quanta (neglecting the magnetic forces on protons).}
\begin{tabular}{c c c c c c c c}
$B_{12}$ &$2E_a$ (eV) &$m_1,m_2$ & $E_m$ (eV) &$Z_o~(a_o)$ &$D_m$ (eV)
&$\hbar \omega_{\parallel}$ (eV) &$\hbar \omega_{\perp 0}$ (eV)\\
\hline
$0.1$ &$-152.8$ &$0,1$ &$-161$ &$0.52$ &$31.7$ &$3.0$ &$2.6$\\
$0.5$ &$-260.4$ &$0,1$ &$-291$ &$0.30$ &$67.5$ &$7.2$ &$8.7$\\
      &		&$0,2$ &$-264$ &$0.32$ &$55.7$ &$6.3$ &\\
$1$   &$-323.0$ &$0,1$ &$-369$ &$0.25$ &$91.0$ &$9.8$ &$14$\\
      &         &$0,2$ &$-337$ &$0.26$ &$76.5$ &$8.8$ &\\
      &		&$0,3$ &$-323$ &$0.26$ &$73.0$ &$8.3$ &\\
$2$   &$-397.0$ &$0,1$ &$-466$ &$0.20$ &$121$  &$13$  &$23$\\
      &         &$0,2$ &$-425$ &$0.21$ &$103$  &$12$  &\\
      &		&$0,3$ &$-408$ &$0.21$ &$98.3$ &$11$  &\\
      &		&$0,4$ &$-398$ &$0.22$ &$96.8$ &$11$  &\\
$5$   &$-514.2$ &$0,1$ &$-623$ &$0.15$ &$173$  &$19$  &$42$\\
      &		&$0,2$ &$-573$ &$0.16$ &$150$  &$18$  &\\
      & 	&$0,3$ &$-550$ &$0.16$ &$143$  &$17$  &\\
      &		&$0,4$ &$-537$ &$0.16$ &$142$  &$16$  &\\
      &		&$0,5$ &$-527$ &$0.17$ &$141$  &$16$  &\\
      &		&$0,6$ &$-519$ &$0.17$ &$140$  &$16$  &\\
$10$  &$-619.2$ &$0,1$ &$-769$ &$0.12$ &$224$  &$25$  &$65$\\
      &		&$0,2$ &$-709$ &$0.13$ &$196$  &$23$  &\\
      &		&$0,3$ &$-682$ &$0.13$ &$188$  &$22$  &\\
      &		&$0,4$ &$-666$ &$0.14$ &$185$  &$22$  &\\
      &		&$0,5$ &$-654$ &$0.14$ &$183$  &$21$  &\\
      &		&$0,6$ &$-645$ &$0.14$ &$183$  &$21$  &\\
      &		&$0,7$ &$-638$ &$0.14$ &$182$  &$21$  &\\
      &		&$0,8$ &$-632$ &$0.14$ &$182$  &$20$  &\\
      &		&$0,9$ &$-627$ &$0.14$ &$182$  &$20$  &\\
      &		&$0,10$&$-623$ &$0.14$ &$181$  &$20$  &\\
\end{tabular}
\end{table}

\narrowtext
\begin{table}
\caption{The energy of the weakly-bound state of H$_2$ in which 
the electrons occupy the $(m,\nu)=(0,0)$ and $(0,1)$ orbitals. 
Here $B_{12}=B/(10^{12}~{\rm G})$, $E_m$ is the energy of the 
molecule, $D_\nu=|E_m|-2|E_a|$ is the dissociation energy of the 
level (neglecting the zero-point oscillation energy of the 
protons), and $Z_o$ is the equilibrium interatomic separation 
($a_o$ is the Bohr radius).}
\begin{tabular}{c c c c}
$B_{12}$ &$E_m$ (eV) &$D_\nu$ (eV)&$Z_o~(a_o)$\\
\hline
0.1 & -167 & 14 & 1.5\\
0.5 & -279 & 19 & 1.3\\
1   & -344 & 21 & 1.1\\
2   & -421 & 24 & 0.99\\
5   & -542 & 28 & 0.89\\
10  & -649 & 30 & 0.72\\
\end{tabular}
\end{table}

\mediumtext
\begin{table}
\caption{Bound-state energy levels of H$_2^+$. 
Here $B_{12}=B/(10^{12}~{\rm G})$,
$E_a$ is the ground state energy of H atom, 
$m$ is the quantum number specifying the electronic excitations
of the molecule, $E_m$ is the electronic energy of the molecule,
$Z_o$ is the equilibrium interatomic separation ($a_o$ is the Bohr
radius), $D_m$ is defined by $D_m\equiv E_a(m)-E_m$, 
$\hbar \omega_{\parallel}$ is the aligned vibrational energy quanta,
and $\hbar \omega_{\perp 0}$ is the transverse vibrational energy
quanta (neglecting the magnetic forces on protons).}
\begin{tabular}{c c c c c c c c}
$B_{12}$ &$E_a$ (eV) &$m$ & $E_m$ (eV) &$Z_o (a_B)$ &$D_m$ (eV) 
&$\hbar \omega_{\parallel}$ (eV) &$\hbar \omega_{\perp 0}$ (eV)\\
\hline
$0.1$ &$-76.4$  &$0$ &$-99.9$ &$0.62$ &$23.5$ &$2.0$ &$3.1$\\
$0.5$ &$-130.2$ &$0$ &$-182$  &$0.35$ &$51.8$ &$4.9$ &$9.8$\\
$1$   &$-161.5$ &$0$ &$-232$  &$0.28$ &$70.5$ &$6.6$ &$16$\\
      &         &$1$ &$-162$  &$0.40$ &$44.8$ &$4.4$ &\\
$2$   &$-198.5$ &$0$ &$-293$  &$0.23$ &$94.6$ &$9.0$ &$25$\\
      &         &$1$ &$-207$  &$0.32$ &$61.5$ &$5.9$ &\\
$5$   &$-257.1$ &$0$ &$-393$  &$0.18$ &$136$  &$13$  &$45$\\
      &		&$1$ &$-284$  &$0.24$ &$91.1$ &$8.6$ &\\
$10$  &$-309.6$ &$0$ &$-486$  &$0.15$ &$176$  &$17$  &$70$\\
      &		&$1$ &$-356$  &$0.19$ &$121$  &$12$  &\\
\end{tabular}
\end{table}

\narrowtext
\begin{table}
\caption{The dissociation energy of H$_2$ molecule 
in the ground state in a superstrong magnetic field.
$D^{(\infty)}$ is the dissociation energy assuming infinite proton
mass, while $D$ includes the (approximate) correction of the molecular
zero-point energy (Eq.~[5.7]). $B_{12}=B/(10^{12}~{\rm G})$,
$\hbar \omega_{\parallel}/2$ is the zero-point energy for the 
aligned vibration, and $\hbar \omega_{\perp}$ is the zero-point energy
for the transverse vibration (Eq.~[5.3]). Note that for 
$B_{12}<0.2$, the ground state is actually the ``weakly-bound''
state (see Sec.~II.A and IV.A), and the zero-point energy has been
neglected. All energies are expressed in eV.}
\begin{tabular}{c c c c c c c c}
$B_{12}$ &$0.1$ & 0.5 &$1$  &$5$  &$10$  &$100$  &$500$\\
\hline
$D^{(\infty)}$ &$14$ & $31$ &$46$ &$109$ &$150$ &$378$ &$615$\\
$D$            &$\simeq 14$ & $21$ &$32$ &$79$  &$110$ &$311$ &$523$\\
$\hbar\omega_\parallel/2$ &$$ & $3.6$&$4.9$ &$9.5$  &$12$ &$22$
		&$31$\\
$\hbar\omega_\perp$     &$$&$6.1$&$9.1$ &$21$  &$28$ &$45$  &$61$\\
\end{tabular}
\end{table}

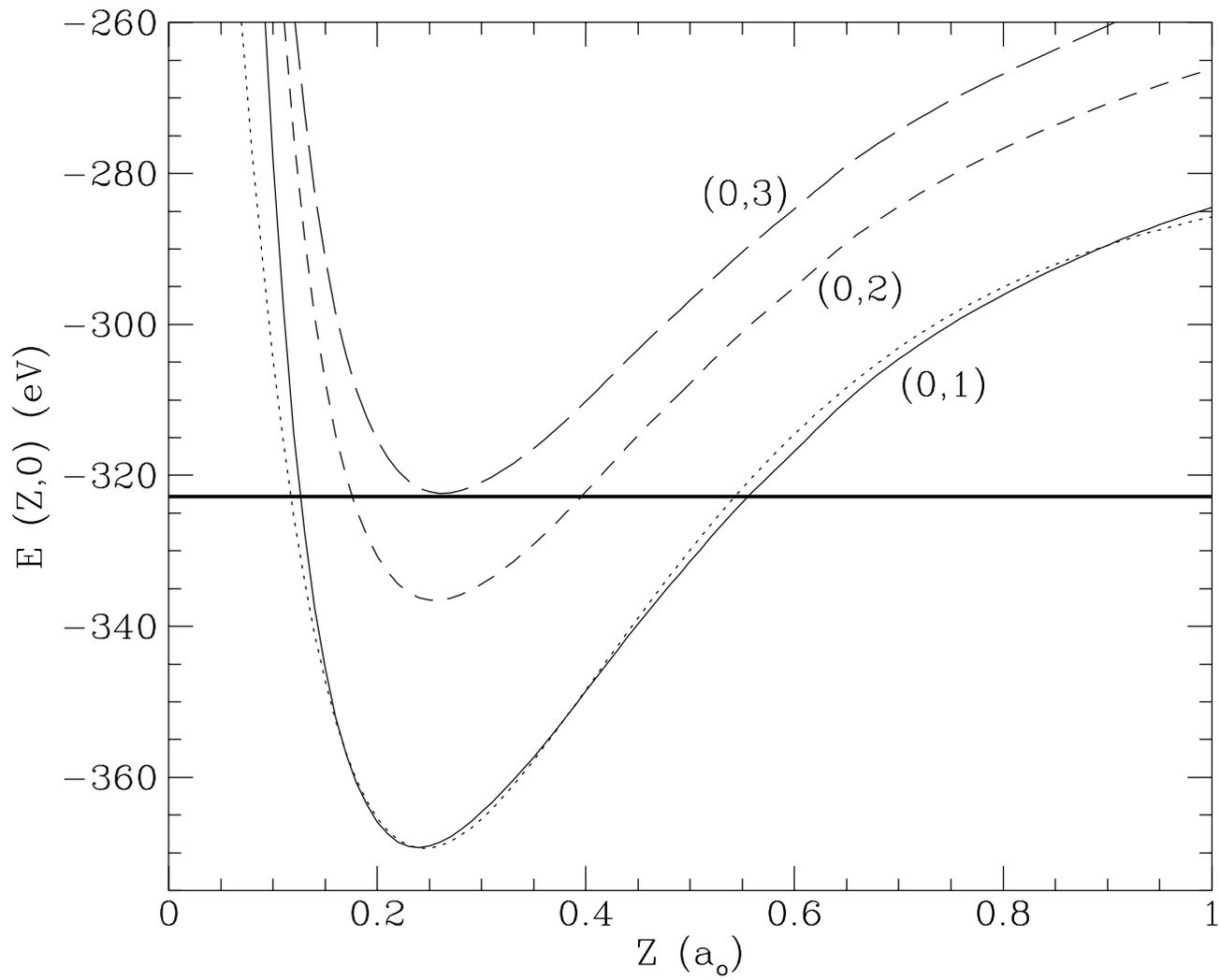
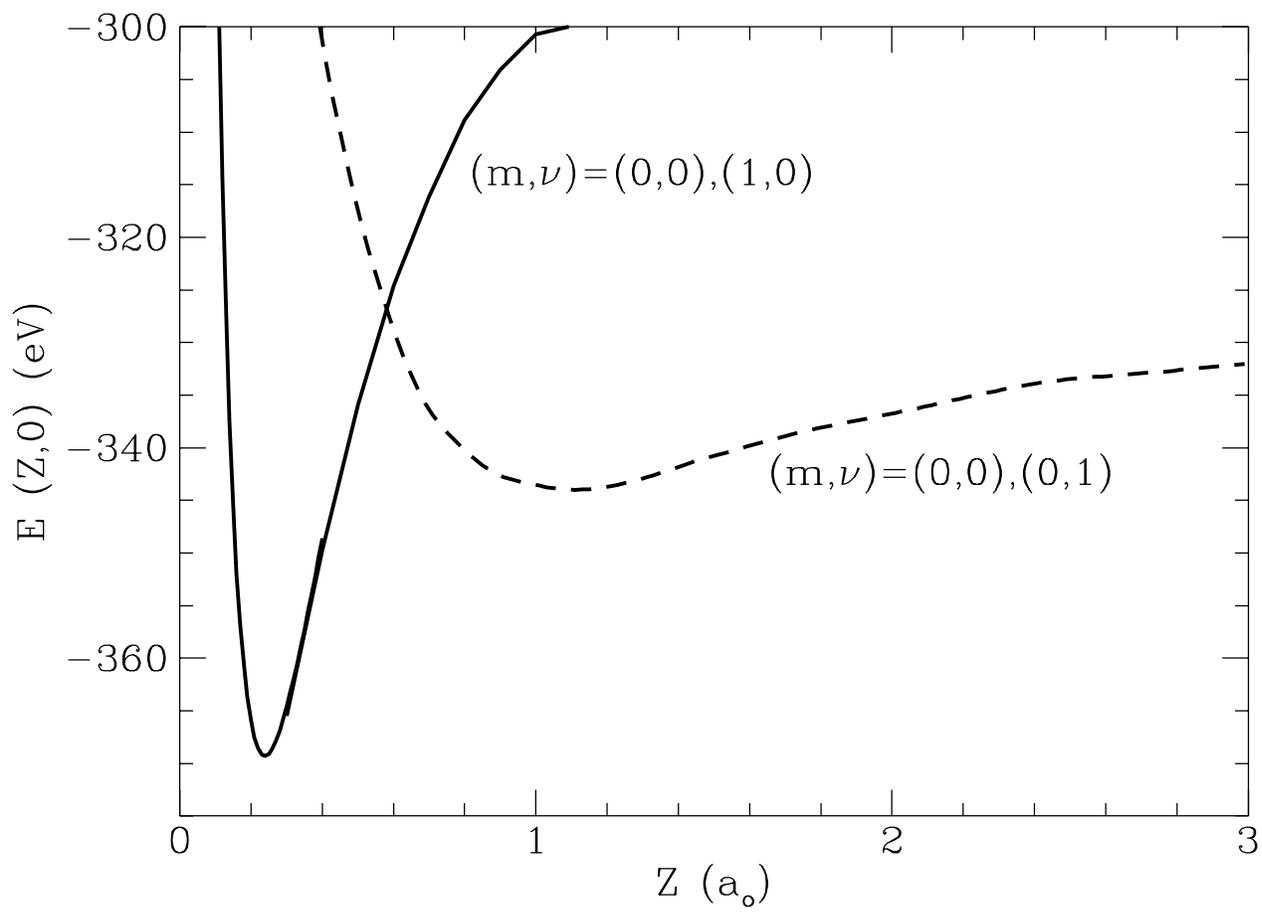
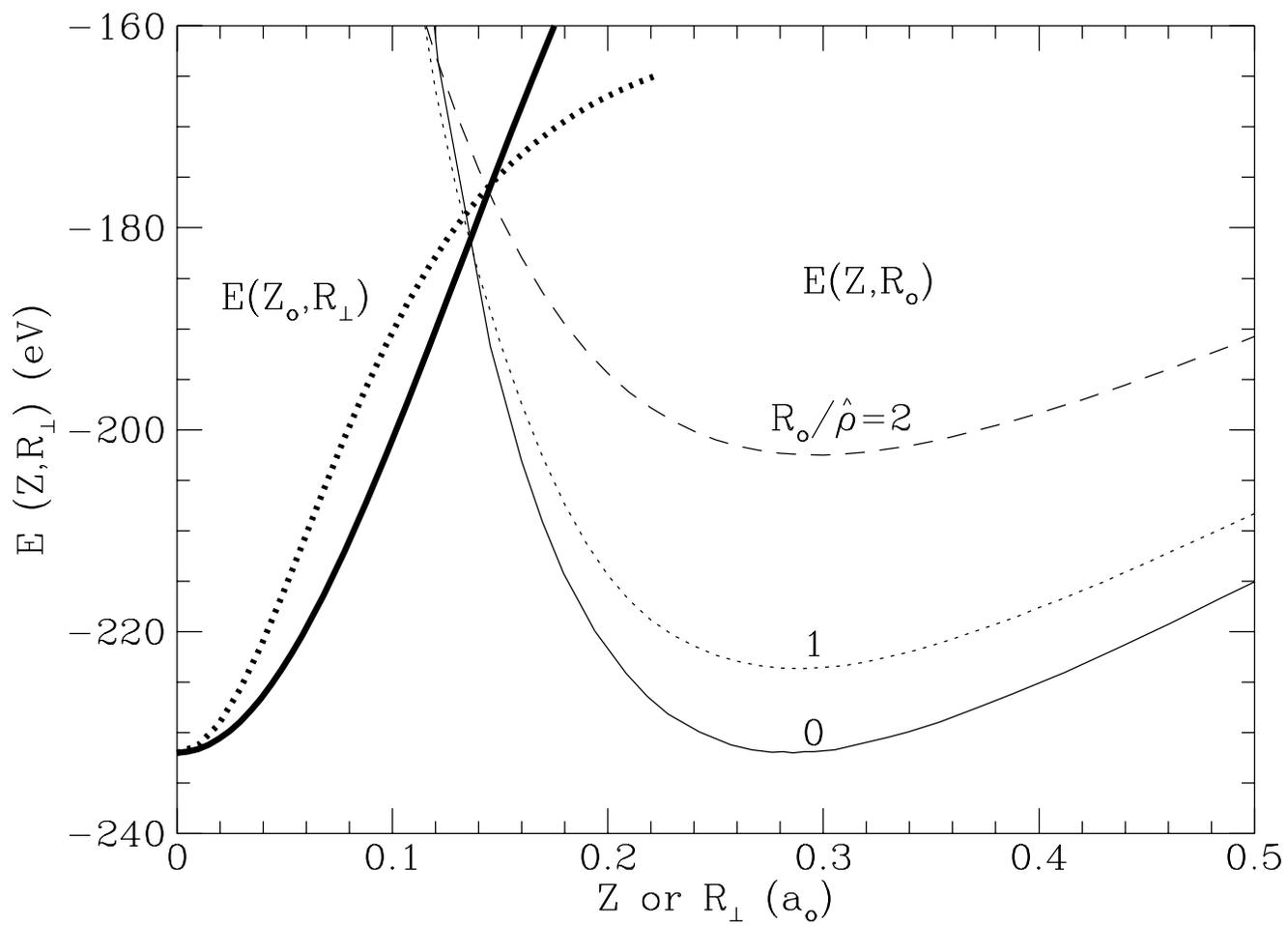
\end{document}